\documentclass[12pt]{article}

\usepackage{amsmath}
\usepackage{amssymb}
\usepackage{amsfonts}
\usepackage{eucal}
\usepackage{wasysym}

\newcommand{\beal}{\begin{subequations} \begin{align}}
\newcommand{\eeal}{\end{align} \end{subequations}}
\newcommand{\be}{\begin{equation}}
\newcommand{\bea}{\begin{eqnarray}}
\newcommand{\eea}{\end{eqnarray}}
\newcommand{\ba}{\begin{array}}
\newcommand{\ea}{\end{array}}
\newcommand{\ee}{\end{equation}}

\makeatletter
\@addtoreset{equation}{section}
\renewcommand{\theequation}{\thesection.\arabic{equation}}
\makeatother
\expandafter\ifx\csname mathbbm\endcsname\relax

\else

\fi
\def\tX{\tilde{X}}
\def\tx{\tilde{x}}
\def\tP{\tilde{P}}
\def\tp{\tilde{p}}
\def\tomega{\tilde{\omega}}
\def\talpha{\tilde{\alpha}}
\def\tbeta{\tilde{\beta}}

\def\pp{pp-wave}
\def\pl{plane-wave}
\def\para{parallelizable}
\def\party{parallelizability}
\def\ppp{parallelizable pp-wave}
\def\susy{supersymmetry}
\def\susys{supersymmetries}
\def\sugra{supergravity}

\def\eom{equations of motion}
\def\christ{Christoffel}
\def\hplus{H_+ \mspace{-25mu} \diagup \mspace{10mu}}

\begin{document}
\baselineskip 18pt

\begin{titlepage} 
\hfill
\vbox{
    \halign{#\hfil         \cr
           SU-ITP-03/05\cr
           SLAC-PUB-9713\cr
           hep-th/0304169 \cr
           } 
      }  
\vspace*{6mm}
\begin{center}  
{\Large {\bf 
String Theory on Parallelizable PP-Waves 
}\\}
\vspace*{15mm}
\vspace*{1mm}

{\bf Darius Sadri${}^{1,2}$ and  M. M. Sheikh-Jabbari$^{1}$}

\vspace*{0.4cm}
{\it ${}^1${Department of Physics, Stanford University\\
382 via Pueblo Mall, Stanford CA 94305-4060, USA}}

\vspace*{0.2cm}

{\it ${}^2${Stanford Linear Accelerator Center,
Stanford  CA 94309, USA }}\\

\vspace*{0.4cm}

{\tt darius@itp.stanford.edu,$\,$jabbari@itp.stanford.edu}

\vspace*{1cm}
\end{center}

\begin{center}
{\bf\large Abstract}
\end{center}
{
The most general parallelizable pp-wave backgrounds which are non-dilatonic
solutions in the NS-NS sector of type IIA and IIB string theories are
considered.
We demonstrate that parallelizable pp-wave backgrounds are necessarily
homogeneous plane-waves, and that a large class of homogeneous plane-waves
are parallelizable, stating the necessary conditions.
Such plane-waves can be classified according to the number
of preserved supersymmetries. In type IIA, these include 
backgrounds
preserving 16, 18, 20, 22 and 24 supercharges, while in the IIB case they 
preserve
16, 20, 24 or 28 supercharges. An intriguing property of parallelizable pp-wave
backgrounds is that the bosonic part of these solutions are invariant 
under T-duality, while the number of supercharges might change under 
T-duality. Due to their
$\alpha^\prime$ exactness, they provide interesting backgrounds for
studying string theory. Quantization of string modes, their 
compactification and behaviour under T-duality are studied.
In addition, we consider BPS $Dp$-branes, and show that 
these $Dp$-branes can be classified in terms of the locations
of their world volumes with respect to 
the background $H$-field.
}

\end{titlepage}

\vskip 1cm

\section{Introduction}
More than a decade ago it was argued that  \pp s provide us with 
solutions of supergravity which are $\alpha'$-exact \cite{Horowitz}.
It has also been shown that any solution of classical general relativity 
(GR) in a special limit known as the Penrose limit, generates a plane-wave 
geometry 
\cite{Penrose}. Plane-waves are a sub-class of \pp s with planar
symmetries. The process of taking the Penrose limit consists of finding a 
light-like geodesic, expanding the metric about that geodesic and scaling 
all the coordinates corresponding to the other directions properly
\cite{Penrose, Veronica1}. The Penrose limit 
has also been extended to supergravity solutions \cite{Guven, Blau}.
However, not much attention was focused on these solutions
until very recently, when it was realized 
that the string theory sigma model on a large class of \pl s is solvable 
\cite{Metsaev}.
The recent interest has been further boosted  by the work of 
Berenstein-Maldacena-Nastase, BMN,
\cite{BMN}, where they noted that a specific \pl\ solution in the type IIB 
supergravity appears as the Penrose limit of $AdS_5\times S^5$ geometry 
and hence is a maximally supersymmetric solution of type IIB theory \cite{Blau}. Based 
on this observation they proposed a correspondence between certain 
operators of  ${\cal N}=4,\ D=4$ super Yang-Mills theory and string theory 
on \pl s. The BMN proposal has by now passed many crucial tests
(see for e.g. \cite{G7}, and references therein),
and has been extended to many other cases with various numbers of
supercharges \cite{Sunil}. 

It has also been argued that all ten dimensional \para\ 
solutions of type IIA, IIB or heterotic supergravities 
are also $\alpha'$-exact \cite{Tseytlin}. In the physics sense, \para\ 
manifolds are 
those for which  there exists a (parallelizing) torsion which makes the 
manifold flat.\footnote{More precise definition of \party\ would be 
presented in section 2.} Besides flat space which is 
trivially \para , generically, \para\ spaces are endowed with torsion 
and hence are usually ignored in the context of classical GR. 
However, they arise naturally in most string theories (with the exception of type I)
where the field strength of the NS-NS two-form field 
$B_{\mu\nu}$ is interpreted as torsion in the target space.

Generally, in these \para\ supergravity solutions only the metric and 
$B_{\mu\nu}$ fields are turned on, i.e., these are non-dilatonic solutions 
in the NS-NS sector of IIA or IIB supergravities.   
As we will prove in section 3, as a result of \party,  the \susy\ 
variation of the gravitinos vanishes for 32 independent solutions 
and hence all the restrictions on the number of supercharges
come from the dilatino variations.

In this work we will focus on \pp s which are \para\ and prove that in 
general the \ppp s are necessarily of the form of homogeneous  \pl s and 
also show that a large class of homogeneous \pl\ geometries are \para , including  
the geometries coming as Penrose limits of $AdS_5\times S^5$ and 
$AdS_3\times S^3$. We also argue that \party\ survives the Penrose 
limit, and hence the Penrose limits of $AdS_{3}\times S^{7}$ geometries\footnote{
Note that this geometry is not a solution of \sugra , this will 
be clarified more in section 3.2.} are also 
\ppp s. We show that general \ppp s form a family of $\alpha'$-exact 
supergravity solutions determined by four real parameters. Then we proceed 
with counting the number of supercharges for these \ppp s. We 
show that the maximum number of supersymmetries for a \ppp\ is 28 which 
corresponds to a IIB \pl\ background with $U(4)$ rotational symmetry,
a solution which has appeared previously in reference \cite{Bena}. 

Performing a Michelson transformation \cite{Michelson}, we study 
toroidal compactification of our solutions and their behaviour under 
T-duality. A remarkable property of \ppp s is that the bosonic part of such 
solutions are invariant under T-duality; however, due to the change in the 
boundary conditions, the number supersymmetries of such T-dual solutions 
might be different. We recall that this property is a generic feature of 
all 
homogeneous \pl s, regardless of whether they are \para\ or not. 
In general, we show that a \ppp\ in type IIA with $N_A$ 
supercharges is T-dual to the same geometry in type IIB with 
$N_B=16$ or $N_B=2N_A-16$ supersymmetries. In addition, there exist
$N_A=20, 24$ solutions for which $N_A=N_B$. Therefore, for $N_A=16,20,24$
there is the possibility of finding self T-dual solutions.

In section 6, we formulate string theory on the most general \ppp \ 
background and show that in light-cone gauge it is solvable with a 
particularly simple spectrum. We also study T-duality at the level of 
string theory. In our case, as a result of the existence of a non-zero 
NS-NS H-field, the right and left-movers of closed string modes enter
differently. Specifically, only left (or right) movers appear in the zero-modes.
This in particular leads to the peculiar property that the 
zero-mode in one direction behaves as the momentum mode of another 
direction. Hence, the
string is probing a non-commutative cylinder where its fuzziness is 
inversely proportional 
to the $B$-field strength, as well as the light-cone momentum. 
We then turn to the question of $D$-branes in the \ppp\ backgrounds 
and briefly discuss which classes of BPS $Dp$-branes can exist in these backgrounds.
A full analysis of these $D$-branes is postponed to future works.

The paper is organized as follows: Section 2 contains a review of
needed material. We first review the definition of \pp s, \pl s and 
homogeneous \pl s and fix our conventions and notations. We then state 
the definition of \party\ and some facts about \para\ manifolds. In 
section 3, we present implications of \party\ for supergravity and its 
solutions. In section 4, we prove that all \para\ \pp s are
homogeneous \pl s and classify the \ppp s by the amount of supersymmetry 
they preserve. In section 5, we discuss toroidal compactification of \ppp 
s as well as their {\it invariance} under a large class of T-dualities.
In section 6, we study superstring theory on \ppp s and work out the 
bosonic and fermionic  string spectrum (in the light-cone gauge). In 
section 7, we study T-duality on the string spectrum. In section 8,
we briefly discuss $Dp$-branes on \ppp s. The last section is devoted 
to concluding remarks and open questions.

\section{Reviews}

In this section we review, very briefly, some of the necessary facts about 
the two basic ingredients of this paper, \pp 
s and \party . For more detailed discussions of these topics the reader is 
referred to the literature \cite{Veronica2, Duff:hr}.

\subsection{Review of pp-waves}

A general class of space-times with interesting properties are given by
{\it pp-waves}. They are defined as space-times which support a 
covariantly constant null Killing
vector field $v^\mu$,
\begin{equation}\label{covariant}
\nabla_\mu v_\nu \: = \: 0\ ,\ \ \ \ \ \  v^\mu v_\mu \: = \: 0\ .
\end{equation}
In the most general form, they have metrics which can be written as 
\begin{equation}\label{ppmetric}
ds^2 \: = \:
-2 du dv -F(u,x^i) du^2 + 2A_j(u,x^i) du dx^j +g_{jk}(u,x^i) dx^j dx^k\ , 
\end{equation}
where $g_{jk}(u,x^i)$ is the metric on the space transverse to a pair of 
light-cone directions given by $u,v$ and the coefficients
$F(u,x^i)$, $A_j(u,x^i)$ and $g_{jk}(u,x^i)$ are constrained by 
(super-)gravity equations of motion.
The pp-wave metric (\ref{ppmetric}) has a null Killing vector
given by $\frac{\partial}{\partial v}$ which is in fact 
covariantly constant by virtue of the vanishing of $\Gamma^v_{v u}$.
The most useful pp-waves, and the ones generally considered in the 
literature, have $A_j = 0$ and are flat in the transverse directions, i.e.
$g_{ij}= \delta_{ij}$, for which the metric becomes
\be\label{pp'metric}
ds^2 \: = \:
-2 du dv -F(u,x^i) du^2 + \delta_{ij} dx^i dx^j\ . 
\ee

Existence of a covariantly null Killing vector field of the space-time 
implies that all the higher dimensional operators built from curvature 
invariants vanish and hence there are no $\alpha'$-corrections to \pp 
s of the form (\ref{pp'metric}) which are solutions of classical 
supergravity \cite{Horowitz}.\footnote{
In general, if $g_{ij}(u,x^i) \neq \delta_{ij}$ the statement
about $\alpha'$-exactness is not true and the transverse metric, $g_{ij}$, 
itself may receive $\alpha'$-corrections, however, there are no extra 
corrections due to the wave part of the metric \cite{Simeon}.} 
For string theory, however, the existence of this null Killing vector
leads to a definition of frequency (in light-cone gauge) which is 
conserved and as a result, the usual problem of non-flat space-times, 
namely the particle (string) creation, is not present.

A more restricted class of pp-waves, the {\it plane-waves}, are those 
admitting a {\it globally defined} covariantly constant null Killing 
vector field. One can show that for \pl s
$F(u,x^i)$ is quadratic in the $x^i$ coordinates of the transverse space,
but still can depend on the coordinate $u$,
$F(u,x^i) = f_{ij}(u) x^i x^j$, so that the metric takes the form
\begin{equation} \label{plane-wave-metric}
ds^2 \: = \:
-2 du dv -f_{ij}(u) x^i x^j du^2 + \delta_{ij} dx^i dx^j \ . 
\end{equation}
Here $f_{ij}$ is symmetric and by virtue of the only non-trivial
condition coming form the equations of motion, its trace is related to
the other field strengths present.
For the case of vacuum Einstein equations, it is traceless.

There is yet a more restricted class of plane-waves, homogeneous plane 
waves, for which $f_{ij}(u)$ is a constant, hence their metric is of the 
form
\be\label{homometric}
ds^2 \: = \:
-2 du dv -\mu_{ij} x^i x^j du^2 + dx^i dx^i \ ,
\end{equation}
with $\mu_{ij}$ being a constant.\footnote{
This usage of the term homogeneous is not universal. For example,
the term symmetric plane-wave has been used in \cite{Blau:2002js}
for this form of the metric, reserving homogeneous for a wider subclass of plane-waves.
}

\subsection{Parallelizability}

An $n$ dimensional manifold M is said to be parallelizable if there exists
a smooth section of the frame bundle, or equivalently, if there exist $n$
smooth sections of the tangent bundle T(M), such that they are linearly
independent at each point of M.  More intuitively, 
a manifold is 
parallelizable if one can cover the whole manifold with a single non-degenerate
coordinate system. 
In general, most manifolds are not
parallelizable.  However, group manifolds are always parallelizable. A
well known result of K-theory due to Adams \cite{Adams} is the
classification of all parallelizable spheres. These consist only of $S^1$,
$S^3$ and $S^7$. One can demonstrate the parallelizability of $S^1$ and
$S^3$ by noting that they are group manifolds: $S^1$ is the group manifold
of $U(1)$, while $S^3$ is the group manifold of $SU(2)$.
However, $S^7$ which can be thought of as the octonions of unit norm,  
is not a group manifold
because the octonionic algebra is non-associative and so the associativity 
property of groups is not satisfied.
This result on parallelizable 
spheres is
closely linked to the Hurwitz theorem \cite{Okubo}, which gives a complete
classification of the unital composition algebras (i.e. Hurwitz algebras),
as the reals, the complex numbers, the quaternions and the octonions (or 
Cayley numbers),
and the fact that $S^1$, $S^3$ and $S^7$ are topologically equivalent to 
the complex numbers, 
quaternions and octonions of unit norm, respectively. The 
non-parallelizability of $S^2$ is
enshrined in the famous no hair theorem \cite{MTW}.
We should also note that, for the reason stated above,  the 
Lorentzian version of $S^3$, i.e., $AdS_3$ 
is also \para .  

There is another definition of parallelizable manifolds due to 
Cartan-Schouten \cite{Cartan-Schouten}:
A manifold is called parallelizable if there exists a torsion which 
``flattens'' the manifold, i.e. makes the Riemann curvature tensor vanish.
We caution the reader that the definitions of \party\ we have given 
here, which is the one assumed in the physics literature, is known in the 
mathematics literature as absolute parallelism, and is a stronger condition than 
the mathematical definition.
In general, absolute parallelism implies parallelism, but not vice-versa.

Let us make explicit the decomposition of the 
connection into a \christ\ piece and a torsion contribution:
\be
\hat{\Gamma}^{\lambda}_{\mu\nu}= {\Gamma}^{\lambda}_{\mu\nu}+ 
T^{\lambda}_{\mu\nu} \ ,
\ee
where ${\Gamma}^{\lambda}_{\mu\nu}$ is symmetric in $\mu\nu$ indices and  
$T^{\lambda}_{\mu\nu}$ (torsion) is anti-symmetric. The curvature 
$\hat{R}_{\mu\nu\alpha\beta}$ may be decomposed in a similar way, into a piece
which comes only from the \christ\ connection and the torsional 
contributions:
\be\label{curvature}
\hat{R}_{\mu\nu\alpha\beta} \equiv {R}_{\mu\nu\alpha\beta}+
\nabla_{\alpha}T_{\mu\nu\beta}-\nabla_{\beta}T_{\mu\nu\alpha}
+T_{\mu\beta\rho}T^{\rho}_{\nu\alpha}-T_{\mu\alpha\rho}T^{\rho}_{\nu\beta}\ .
\ee
The \party\ condition then simply becomes 
$\hat{R}_{\mu\nu\alpha\beta}=0$.  
If
the modified Ricci tensor
\be\label{Riccipara}
\hat{R}_{\mu\nu} \equiv {R}_{\mu\nu}+\nabla_{\alpha}T^{\alpha}_{\mu\nu}
-T_{\mu\lambda\rho}T^{\lambda\rho}_{\nu}
\ee
is zero, the manifold is said to be Ricci-\para . Note that the generalized Ricci 
tensor $\hat{R}_{\mu\nu}$ is not symmetric in its $\mu\nu$ indices.
For a manifold to be Ricci-\para , the symmetric and anti-symmetric parts 
of 
$\hat{R}_{\mu\nu}$ should both vanish, namely 
$\nabla_{\alpha}T^{\alpha}_{\mu\nu}=0$ and 
${R}_{\mu\nu}-T_{\mu\lambda\rho}T^{\lambda\rho}_{\nu} =0$.  

With the above definitions, the parallelizing torsion (in an orthonormal frame)
for a group manifold is given by the structure constants of the group 
algebra \cite{Cartan-Schouten}. 

\section{Parallelizability and supergravity}
\label{pandsusy}

Since in most string theories, a torsion field naturally arises,
one may look for implications of \party\ for supergravities and 
their solutions. Here we mainly focus on type II theories, however, most 
of our arguments can be used for heterotic theories as well.

First we recall that the NS-NS part of the \sugra\ action is of the form
\be\label{sugraaction}
S=\frac{1}{l_p^8} \int d^{10}x \sqrt{-g} e^{-2\phi}\left( R+4
(\nabla_\mu\phi)^2 -\frac{1}{12} H^2\right)\ .
\ee
Since we are interested in solutions involving only the metric and 
torsion, we set the dilaton field $\phi$ to a constant. Then the \sugra\ 
equations of motion for the metric and $B_{\mu\nu}$ field are
\begin{subequations}
\label{sugra:eom}
\begin{align}
R_{\mu\nu}-\frac{1}{4} H_{\mu\rho\lambda}H_{\nu}^{\rho\lambda}=0\ , \\
\nabla_{\alpha}(\sqrt{-g} H^{\alpha}_{\mu\nu})=0, \ \ \ \ H=dB\ .
\end{align}
\end{subequations}
If we define $\frac{1}{2}H_{\alpha\mu\nu}$ as torsion $T_{\alpha\mu\nu}$,
as we discussed in previous section, the \sugra\ equations for the metric 
and $B_{\mu\nu}$ fields are nothing but the Ricci-\party\ condition. 
Hence all \para\ manifolds (which are obviously also Ricci-\para ) 
satisfying the constant dilaton constraint \cite{PandoZayas}
\be\label{Dilaton}
R=\frac{1}{12} H^2
\ee
and have a closed torsion (i.e. $dH=0$), are solutions of \sugra .

For any \sugra\ solution one may wonder about  $g_s$ and $\alpha'$ 
exactness as well as (classical) stability. The \para\ solutions which we 
are interested in are non-dilatonic and hence they are $g_s$ independent. 
The $\alpha'$-exacctness of these \sugra\ solutions were studied long 
ago in reference \cite{Tseytlin}. Computing the second $\alpha'$ contributions 
to the string theory $\beta$-functions, it was shown that such 
contributions are zero for \para\ manifolds.\footnote{Of course, as 
in any field theory with more than one coupling, the two loop 
renormalizations are scheme dependent, and according to reference 
\cite{Tseytlin} in a specific scheme such contributions are zero for 
\para\ manifolds. For further details the reader is referred to that 
reference.} Therefore the \para\ solutions of \sugra\ are exact up to 
order $\alpha'^2$. It has been argued that this property is expected to 
remain to all orders in $\alpha'$ \cite{Tseytlin}.

\subsection{Parallelizability and supersymmetry}

In this section we will demonstrate one of the interesting implications of 
\party\ for supergravity solutions. Here we restrict 
ourselves to type II theories. Our conventions can be found in the appendix.

{\it \bf Theorem:}

If we denote the \susy\ variations of the gravitinos by 
$\delta\psi^\alpha_{\mu},\ \alpha=1,2$, 
\party\ implies that $\delta\psi^\alpha_{\mu}=0$ has the maximal number 
of solutions (32) and thus does not lead to any restrictions on the number of 
supercharges. Therefore all the \susy\  restricting conditions come from 
the dilatino variations.

To prove the above theorem, we make use of the \susy\ variations for type
IIA and IIB string theories, whose complete expression is presented in the 
appendix.
For our \para\ backgrounds, which do not depend on RR fields,
the IIA and IIB expressions both reduce to
\be\label{Dhat}
\hat{\mathcal{D}}_{\mu} \: = \: \partial_{\mu}
+\frac{1}{4} {\hat \omega}_\mu \ ,
\ee
where ${\hat \omega}_\mu$ is the torsional spin connection\footnote{
In string theory terminology $\omega_\mu^{ab}$ only captures 
the torsion independent part of the connection and is completely
determined by the \christ\ connection.
}
and is given by
\be\label{connection}
{\hat \omega}_\mu \: = \:
\omega_\mu^{ab} \Gamma_{ab}+\frac{1}{2} 
\sigma_3 \Gamma^{ab} H_{\mu ab} \ .
\ee

The Killing spinor equation reads
\begin{equation}\label{Killingspinor}
\delta\psi_{\mu}= \hat{\mathcal{D}}_\mu \epsilon \: = \: 0 \ .
\end{equation}
The commutator of two covariant derivatives acting on a spinor $\psi$
contains a contribution from the curvature and also a torsion piece
multiplying a covariant derivative
\begin{equation} \label{}
  [ \hat{\mathcal{D}}_\mu , \hat{\mathcal{D}}_\nu ] \: \epsilon \: = \:
  \hat{R}_{\mu\nu ab} \Gamma^{ab} \epsilon \:  \:
  - \: T^\lambda_{\mu\nu} \: \hat{\mathcal{D}}_\lambda \: \epsilon \ .
\end{equation}
Now we note that for \para\ manifolds, by virtue of the vanishing 
of the
curvature computed from the connection with parallelizing torsion, 
the Killing spinor equation implies
\begin{equation} \label{integrability}
  [ \hat{\mathcal{D}}_\mu , \hat{\mathcal{D}}_\nu ] \: \epsilon \: = \: 
- T^\lambda_{\mu\nu} \: \hat{\mathcal{D}}_\lambda \: \epsilon \ .
\end{equation}
The left-hand side is zero for solutions of 
(\ref{Killingspinor}). That is, (\ref{integrability}) is an 
integrability 
condition on the differential equation (\ref{Killingspinor}).
We find that the integrability condition imposes no extra constraints on the solution
(this is special to parallelizable manifolds). Note that the vanishing 
of the Riemann curvature tensor is central to this argument; Ricci parallelizability, i.e.,
the vanishing of the
Ricci tensor computed with parallelizing torsion, does not suffice.
Since the Killing spinor equation is a first order differential equation,
it can be solved by introducing the ansatz,
\begin{equation} \label{ansatz}
  \epsilon(x) \: = \: {\cal W}(x)
   \: \chi \ ,
\end{equation}
where
\begin{equation} \label{ghat}
  {\cal W}(x) \: = \: \mathcal{P} e^{-\frac{1}{4}\int^x \hat\omega \cdot dl} \ ,
\end{equation}
and $\chi$ is a 32 component spinor\footnote{
This has an interpretation as a gauge transformation, where the Killing spinors are pure gauge
when the connection is taken to be that with parallelizing torsion.}
and $\mathcal{P}$ denotes the path ordering symbol.
Plugging the ansatz \eqref{ansatz} into (\ref{Killingspinor}), the 
gravitino Killing equation reduces to 
\be\label{chi}
\partial_{\mu} \chi =0\ .
\ee
That is, equation (\ref{Killingspinor}) is solved for any constant $\chi$.
This provides $32$ independent solutions for both types IIA and IIB.

So far we have shown that for any parallelizable manifold, the Killing spinor equation 
arising from vanishing of the gravitino variation is satisfied for any spinor of the form 
(\ref{ansatz},\ref{chi}) and the correct number of unbroken 
supersymmetries is only determined with the zero
dilatino \susy\ variation condition.\footnote{
This is in contrast to the usual situation with plane-waves, where the 
dilatino equation is redundant and the supersymmetries are determined 
only by the gravitino equation.}
For a constant dilation background with the only non-vanishing flux 
the NS-NS field strength, in string frame, this condition is
\begin{equation}
\label{dilatino:variation}
  \delta \lambda \: = \:
  - \frac{1}{4} \Gamma^{abc} H_{abc} \sigma^3  \epsilon \: = \: 0 \ .
\end{equation}
Therefore, the associated condition for the existence of Killing spinors is
$H_{abc} \Gamma^{abc} \: \epsilon^\alpha \: = \: 0$, for $\alpha=1,2$.

\subsection{Some examples of ten dimensional \para\ geometries}

Direct products of \para\ manifolds are also \para.
As a 
famous example we mention $AdS_3\times S^3 \times M_4$ 
\cite{Horowitz:1996ay,Maldacena:1997re}. The $AdS_3\times S^3$ is \para\ 
by virtue of being a group manifold. However, \sugra\ \eom\  
for the dilaton, namely equation (\ref{Dilaton}), forces the $AdS$ and sphere to have 
the same radii. 
Therefore,
this non-dilatonic solution is \para\ if $M_4$ 
is \para , and since the only \para\ four dimensional manifold
which is compatible with \eqref{Dilaton},
is flat 
space, it is \para\ if $M_4$ is $R^4$ or $T^4$. However, if $M_4$ is Ricci 
flat,
e.g. $M_4=K3$, this solution is only Ricci-\para .

%
Another explicit example of a parallelizable ten dimensional 
geometry (which is not, however, a \sugra\ solution) background 
can be built out of $AdS_{3}$ and $S^{7}$ with the
following background field configuration
\begin{subequations} \label{Pando:Zayas}
\begin{align} 
  ds^2 \: &= \:
    R_1^2 \left( \frac{du^2}{u^2} \: + \: u^2 \left( -dt^2 + dx^2 \right) \right) \ , \\
  B_{tx} \: &= \: R_1^2 u^2 \ ,
\end{align}
\end{subequations}
for the $AdS_3$ part, and
\be \label{S7}
  ds^2 \: = \:
    \frac{R_2^2}{4} \left(
    d \mu^2 + \frac{1}{4} sin^2 \mu (\sigma_i - \Sigma_i)^2 \: + \:
    \lambda^2 ( \cos^2 \frac{\mu}{2} \sigma_i + \sin^2 \frac{\mu}{2} 
\Sigma_i )^2
    \right) \ ,
\ee
with
\begin{subequations}
\begin{align} \label{}
  \sigma_1 \: &= \: \cos \psi_1 d \theta_1 \: + \: \sin \psi_1 \sin \theta_1 d \phi_1 \ , \\
  \sigma_2 \: &= \: - \sin \psi_1 d \theta_1 \: + \: \cos \psi_1 \sin \theta_1 d \phi_1 \ , \\
  \sigma_3 \: &= \: d \psi_1 \: + \: \cos \theta_1 d \phi_1 \ ,
\end{align}
\end{subequations}
and the $\Sigma_i$ given by the same 
relations as $\sigma_i$, but with
$\psi_1,\theta_1,\phi_1 \rightarrow \psi_2,\theta_2,\phi_2$, e.g. see 
\cite{Englert}.
For $S^7$, $\lambda^2=1.$\footnote{
The round seven-sphere is the coset space $SO(8)/SO(7)$,
with the standard metric induced from an isometric embedding
in $\mathbb{R}^8$, making manifest
its $SO(8)$ isometry.
One may analogously define the squashed seven-sphere, topologically equivalent to
$S^7$, as the distance sphere in the
projective quaternionic plane,
with metric derived from an isometric embedding.
A derivation of this metric is presented in
\cite{Duff:hr}.
To yield an Einstein space, $\lambda^2$ can only take on two discrete
values, with $\lambda^2=1$ corresponding to $S^7$ and $\lambda^2=\frac{1}{5}$ giving
the metric on the squashed $S^7$.
An equivalent description of the squashed seven-sphere can be constructed by noting that
the round seven-sphere can be described as a fiber-bundle, with base $S^4$ and fiber $S^3$.
The squashing arises from a change of the $S^3$ radius over the base.
}
In the above, $R_1$ and $R_2$ are the radii of $AdS_3$ and $S^7$ respectively,
while $\sigma_i$ and $\Sigma_i$ parametrize three-spheres.
$S^7$ is the manifold spanned by unit octonions, where the octonion 
algebra\footnote{A discussion of octonions, their algebra, and their 
appearance as the
parallelizing torsion on $S^7$ can be found in 
\cite{Gursey:1983yq,Gunaydin:1973rs}.}
is
\begin{subequations}\label{octanion}
\begin{align} 
O_A \: O_B \: &= \: -\delta_{AB} \: + \: f_{ABC} O_C \ \ \ \ \ \ \ \
(A=0,a,\hat{a})
 \ , \\
f_{0 a \hat{b}} \: &= \: \delta_{ab} \ , \: \: \: \: \: \: \: \: \: \:
f_{a b c} \: = \: \epsilon_{abc} \ , \: \: \: \: \: \: \: \: \: \:
f_{a \hat{b} \hat{c}} \: = \: - \epsilon_{abc} \ ,
\end{align}
\end{subequations}
where $0$ corresponds to $\mu$ and  the indices ranging over $a,b,c=1,2,3$,  
and $\hat{a},\hat{b},\hat{c}=4,5,6$ correspond to $\sigma_i$ and $\Sigma_i$ 
respectively.
The components of the torsion three form field  for this manifold, 
in an orthonormal frame, are given by the octonionic structure constants as
\be \label{S7:structure}
T_{a b c} \: = \: \frac{2}{R_2} \: f_{abc} \ ,
\ee
and the Ricci tensor and square of the three-form are
\be \label{Ricci:S7}
R_{ab} \: = \: 
T_{ab}^2 \: = \: \frac{6}{R_2^2} \: \delta_{ab} \ .
\ee
Note that the torsion field is {\it not} a closed form\footnote{
We would like to thank Jose Figueroa-O'Farrill for pointing 
this out to us.}, in fact $dT=\frac{1}{2} T_{\alpha\beta\lambda}
T^{\lambda}_{\mu\nu}dx^{\alpha}\wedge 
dx^{\beta}\wedge dx^{\mu}\wedge dx^{\nu}$. 
Hence this parallelizing torsion is {\it not} 
the field strength of any (NS-NS) two form field and consequently 
$AdS_3\times S^7$ geometry discussed here, although being \para , is not a 
\sugra\ solution. We would like to point out that
this non-closedness of torsion is related to the non-associativity of the 
octonion algebra, and for all group manifolds
the parallelizing torsion is a closed form.



The round $S^7$ provides an example of a parallelizable geometry, i.e.,
its modified Riemann tensor vanishes. However, a specific deformation of
the round $S^7$, the squashed $S^7$ ($S^7_q$) with the metric given in
(\ref{S7}) for $\lambda^2=1/5$, leads to a manifold which is only
Ricci-\para \cite{Duff:hr}. The torsion still has the form of 
\eqref{S7:structure},
however, now the Ricci tensor and $R_2$ are related via
$R_{ab}=\frac{54}{5R_2^2} \delta_{ab}$.\footnote{Note that the Ricci
tensor in \eqref{Ricci:S7} is derived from the connection without
torsion.} Then $AdS_3\times S^7_q$ is a ten dimensional Ricci-\para\ 
manifold.

\section{Parallelizable pp-waves}

In this section we construct the most general \pp\ which is also \para . 
Such solutions are ``doubly $\alpha'$-exact'' backgrounds of string theory 
in the sense that they are both \para\ and have a covariantly constant null 
Killing vector field. Here we first prove a theorem showing that all 
\para\ pp-waves are homogeneous \pl s. Then in the second part of this section 
we will classify all the \para\ \pp s by the number of their \susys . In 
the last part of this section, as an example, we work out the Penrose limit 
of $AdS_{3}\times S^{7}$, and show that these solutions only preserve 
16 (kinematical) \susys .

\subsection{Parallelizable pp-waves as homogeneous plane-waves}
\label{ppp:hom}

{\it \bf Theorem:}

All parallelizable pp-waves are homogeneous plane-waves.

To prove this assertion, we begin 
with the most general ten dimensional \pp\ geometry whose metric is given 
in equation (\ref{ppmetric}) where now $i,j=1,2, \cdots, 8$. Of course, the 
functions $F, A_i,g_{ij}$ are chosen to satisfy \sugra\ equations of motion. 
The next step is to write the most general NS-NS $H$ field compatible with 
the covariantly constant Killing vector $v^{\mu}$. This nails down the 
choices for the $H$-field to
\be\label{Hfield}
H_{uij}= h_{ij}(u,x^i)\ ,
\ee
and all the other components zero.
Now, we are ready to impose the \party\ conditions. We may start by
imposing \sugra\ equations of motion. The Ricci curvature for the 
metric (\ref{ppmetric}) may be found in \cite{Gibbons}.
Then we note that the only non-zero $g^{u\mu}$ components of the inverse 
metric is $g^{uv}=-1$, therefore
\be\label{H^2}
H_{\mu\alpha\beta} H_{\nu}^{\ \alpha\beta}=\delta_{u\mu}\delta_{u\nu}
h_{ij}h_{kl}g^{ik}g^{jl}\ ,
\ee
and hence all the components of the Ricci curvature except $R_{uu}$ should be 
zero by virtue of the \sugra\ equations of motion or equivalently the 
Ricci-\para\ conditions. However, first we note that $\hat{R}_{ijkl}=0$
implies that
\be\label{Rij}
R_{ijkl}=0\ \Longrightarrow\ g_{ij}(u,x^i)=g_{ij}(u) \ .
\ee
Before imposing other \sugra\ \eom\ we note that using (\ref{Rij}), after 
a coordinate transformation, the metric can always be brought to the form
\[
ds^2 \: = \:
-2 du dv -\tilde{F}(u,x^i) du^2 + 2\tilde{A_j}(u,x^i) du dx^j +dx^i dx^i\ . 
\]
Then $R_{ui}=0$ leads to
\be
\partial_i F_{ij}=0, \ \ \ \ F_{ij}=\partial_{[i} \tilde{A}_{j]}\ ,
\ee
and $R_{uu}=\frac{1}{4} H_{uij}H_{u}^{\ ij}$ results in
\be\label{Ruu}
\frac{1}{2}\nabla^2\tilde{F}=\frac{1}{4} (F_{ij}F^{ij}+ h_{ij} 
h^{ij})\ .
\ee
Note that the freedom in defining $v$ coordinate leads to a $U(1)$ gauge 
symmetry in the definition of $\tilde A_i$ and equation (\ref{Ruu}) is written 
in the Lorentz gauge, $\partial_i A_i=0$. Finally the equation of motion
for the $H$ field implies that
\be
\partial_i h_{ij}=0\ .
\ee
 
Next we impose the \party\ condition. It is easy to see that the only remaining 
non-trivial equations come from ${\hat R}_{uiuj}=0$ and 
${\hat R}_{uijk}=0$. In order to avoid lengthy calculations we only 
summarize the results:
\bea\label{solution}
h_{ij}(u,x^i)&=&h_{ij}={\rm constant}\ ,   \cr
{\tilde F}&=&\mu_{ij} x^i x^j\ ,\ \  \mu_{ij}={\rm constant}\ ,\cr
{\tilde A}_i&=&\frac{1}{2}F_{ij} x^j\ ,\ \  F_{ij}=-F_{ji}={\rm 
constant}\ 
,
\eea
and
\be
\mu_{ij}=\frac{1}{4}(h_{ik}h_{jk} +F_{ik}F_{jk})\ .
\ee
In summary, the most general \para\ \pp\ is of the form
\bea\label{pppwave1}
 ds^2  &=& -2 du dv -\mu_{ij} x^i x^j du^2 + F_{ij} x^j du dx^j +
dx^i dx^i\ , \cr  
H_{uij} &=& h_{ij}={\rm constant} \ .
\nonumber
\eea
It can be shown that we can still use rotations in the transverse space 
($x^i$ directions) to remove the $dudx^i$ terms in the metric. Note that 
such rotations will not change $H_{uij}$. We will further 
discuss this coordinate transformation in section 5. Therefore, 
the most general \para\ \pp\ can also be written as\footnote{
It may seem that geometries of the form ${ppp-wave}_d\times 
{\cal M}_{10-d}$, where ${\cal M}$ is a non-flat, parallelizable manifold and 
${ppp-wave}_d$ is a \para\ \pp\ of the kind we have discussed, should be 
included in our \ppp\ list. In fact 
since $d$ can only be even, $d\geq 4$, the only possibility 
for ${\cal M}$ is $S^3\times S^3$. Although such geometries are \para , 
they do not satisfy the constant dilaton constraint (\ref{Dilaton}), and hence 
are not non-dilatonic \sugra\ solutions.}
\bea\label{pppwave2}
 ds^2  &=& -2 du dv -\mu_{ij} x^i x^j du^2 +dx^i dx^i \ , \cr
H_{uij} &=& h_{ij}={\rm constant}\ ,
\eea
with
\be\label{muij}
\mu_{ij}=\frac{1}{4} h_{ik}h_{jk}\ .
\ee
This is of the form of a homogeneous \pl , completing the proof.

In general the converse may not hold, i.e., not all
homogeneous \pl\ geometries are \para . In order to see which homogeneous 
\pl s are \para , we note that ${h_{ij}}$ being an anti-symmetric $8\times 
8$ matrix, can always be brought to a block diagonal 
form by O(8) rotations, so that the only non-zero components are
\be\label{foura}
h_{12}=-h_{21}=2a_1\ , \ 
h_{34}=-h_{43}=2a_2\ , \  
h_{56}=-h_{65}=2a_3\ , \  
h_{78}=-h_{87}=2a_4\   .
\ee
Then eq. (\ref{muij}) yields 
\be
\mu_{ij}=diag(a_1^2,a_1^2,a_2^2,a_2^2,a_3^2,a_3^2,a_4^2,a_4^2)\ .
\ee
Therefore, all homogeneous \pl s given by the metric (\ref{homometric}) are 
\para\ if and only if $\mu_{ij}$ has doubly degenerate eigenvalues.

The above theorem can be understood more intuitively by noting that \party , 
by definition, forces the covariantly constant null Killing vector field
of the \pp , in our case $\partial_v$, to be globally defined. Furthermore, 
it forbids the torsion from having any non-trivial space-time dependence; 
leaving us with the homogeneous \pl s as the only possibility.

We see from eq. (\ref{foura}), that in the most general case, a \ppp\
is completely determined by four real numbers, $a_i$. Depending on 
the values of $a_i$'s, the O(8) rotational symmetry of the transverse
space is broken to sub-groups, however, for the generic case (all $a_i$ 
taking different values) there remains a U(1)$^4$ symmetry. In the most 
symmetric case, where all $a_i$'s are equal, we have a U(4) symmetry. 
The case with $a_3=a_4=0$ and $a_1=a_2$ arises from the Penrose 
limit of $AdS_3\times S^3$ \cite{Russo}. The \pl s similar to
those of eq. (\ref{pppwave2}) have also been considered in \cite{Ganor1, 
Ganor2}.

\subsection{Supersymmetry counting and classification}
\label{susy:counting}

%

In section \ref{pandsusy} we showed that, for \para\ backgrounds, the gravitino
variations impose no constraints on the number of supersymmetries; hence, the
number of unbroken supercharges is determined entirely by the variation of the
dilatino \eqref{dilatino:variation}.
We write the two Majorana-Weyl spinors as two Majorana spinors as in the
appendix, subject to the appropriate chirality conditions \eqref{chirality}.

For the backgrounds we are considering, the non-zero spin connection components are
$\omega_+^{\ -i}=\frac{1}{2} \mu^i_{\ j} x^j$.
This together with the relations \eqref{properties} implies that $\hat{\omega}^2=0$,
and hence the expansion of $\cal{W}$ is linear in $\hat{\omega}_\mu$.
We also have $[ {\cal W} , \Gamma^+ \hplus ]=0$.
Therefore, the supersymmetry variation of the dilatino can be reduced to
\be
\label{nice:killing:spinor:eqn}
  \delta \lambda \: = \:
  - \frac{1}{4} \Gamma^{abc} H_{abc} \: \chi \: = \: 0
\ee
for $\chi^\alpha,\alpha=1,2$,
constant Majorana spinors \eqref{ansatz}.

For these backgrounds, the field strength
components with one leg along the light-cone direction $x^+$
are non-vanishing, and there are no purely transverse contributions. As a result, the
dilatino variation always contains a $\Gamma^+$ acting on $\chi^\alpha$.
In the most general \para\ background \eqref{foura},
\eqref{nice:killing:spinor:eqn} imposes as a condition for the existence of Killing spinors the
requirement
\be \label{killing:spinor:final}
\Gamma^{+} \hplus \chi^\alpha \: = \: 0 , 
\ee
with the two spinors subject again to their chirality conditions\footnote{
See \eqref{chirality}.},
and we have defined
$\hplus \equiv H_{+ij} \Gamma^{ij}$.\footnote{
See \eqref{H:contraction} for an explicit realization of $\hplus$.}

To analyze the dilatino variation, it is easiest to work in a specific basis adapted to the
problem, and we choose the basis \eqref{basis}. We can write a general spinor as
\be
\Psi \: = \:
\left(
\begin{matrix}
\phi \\
\omega \\
\xi \\
\lambda
\end{matrix}
\right) \,
\ee
with $\phi,\omega,\xi,\lambda$, all eight component column vectors.
The chirality conditions imply, in this basis,
\be
\label{right:chirality}
\Gamma^{11} \: \Psi_R \: = + \Psi_R \ \Longrightarrow \
\Psi_R \: = \:
\left(
\begin{matrix}
\phi \\
0 \\ 
0 \\ 
\lambda
\end{matrix}
\right) \ \ \ \ \ then \ \ \ \ \
\Gamma^+ \: \Psi_R \: \propto \:
\left(
\begin{matrix}
0 \\
\lambda \\
0 \\
0
\end{matrix}
\right) \ ,
\ee
for right handed spinors, and
\be
\label{left:chirality}
\Gamma^{11} \: \Psi_L \: = - \Psi_L \ \Longrightarrow \
\Psi_L \: = \:
\left(
\begin{matrix}
0 \\ 
\omega \\
\xi \\
0 \\ 
\end{matrix}
\right) \ \ \ \ \ then \ \ \ \ \
\Gamma^+ \: \Psi_L \: \propto \:
\left(
\begin{matrix}
\xi \\
0 \\
0 \\
0
\end{matrix}
\right) \ ,
\ee
for left handed spinors.

To analyze the Killing spinor equation,
start with an arbitrary left-handed spinor $\Psi_L$, of the form \eqref{left:chirality},
then subtract off components along the subspace projected out by $\Gamma^+$
($\omega$ above).
The remaining spinor is of the form
$\tilde\Psi_L =
\left(
0 \
0 \
\xi \
0
\right)
$.
Since $\Gamma^+$ commutes with $\hplus$, we may add to $\tilde\Psi_L$
any arbitrary spinor components
on the subspace projected out by $\Gamma^+$, and these do not contribute to the
Killing spinor equation.
These unconstrained components provides a minimal set of Killing spinors,
eight for each Majorana spinor above, for a total of 16. These are the
standard or kinematical Killing spinors which exist on any \pl,
so are always present after a Penrose-Gueven limit.
The kinematical supercharges are thus in the kernel of $\Gamma^+$.
Their presence is a direct consequence of the existence of
a null Killing vector on the \pp\ space-time manifold.

Write $\hplus$ as diag(A,B,A,B),
with A and B themselves diagonal matrices with eigenvalues that are doubly degenerate
(up to a sign).
Now require $\hplus \tilde\Psi_L=0$. This amounts to the equation $A \xi = 0$.
The entries of $A$ are a set of four independent algebraic functions of
the $a_i$ from which the field strength is constructed \eqref{foura},
given by $c_i,i=1,2,3,4$, in equation \eqref{constants}.
Some of these $c_i$ may vanish identically for a given background, and the components
of $\xi$ acted on by these can be taken to be arbitrary. They provide additional
Killing spinors beyond the standard kinematical ones.
The components of $\xi$ which are not projected out by the action of A in a given
background must then vanish if $\Psi_L$ is to provide a solution of the Killing
spinor equation, and so do not contribute any additional supercharges.\footnote{
\label{foot:label}
Notice that it is not possible, if all $a_i$ are non-zero,
to simultaneously have some $c_i=0,i=1,2,3,4$ and
another $c_j=0,j=5,6,7,8$. If we had not subtracted off
the kernel of $\Gamma^+$ (the kinematical supercharges $\omega$),
then $\hplus \Psi_L = 0$ alone would yield
two simultaneous equations, $A \xi = 0$ and $B \omega = 0$, but
since $\omega$ was assumed arbitrary, $B=0$ and so $c_j=0,j=5,6,7,8$, which is
possible only for all $a_i=0$, contradicting the earlier assumption.
This is another way to see that the presence
of $\Gamma^+$ is central to the existence of any Killing spinors on non-trivial
\ppp s.}
A similar argument carries through for the case of right handed spinors, where
the constraint becomes $B \lambda = 0$, and the four algebraic functions of interest
are now $c_i,i=5,6,7,8$.

As an example, take the most symmetric non-trivial  background,
with $a_1=a_2=a_3=a_4 \ne 0$, possessing
a U(4) symmetry (whose $\mathbb{Z}_4$ center acts by interchanging the four planes).
Let us assume that we have subtracted off the kinematical supercharges.
For a left-handed spinor, the equations whose solutions gives rise to non-kinematical
supercharges, are $c_i=0,i=1,2,3,4$,
with the $c_i$'s given in \eqref{constants}.
Of these, three are identically satisfied and one is not. As
a result, $3/4$ of $\xi$ are unconstrained.
For a right-handed spinor, the equations would be $c_i=0,i=5,6,7,8$,
and of these, none
are identically satisfied, so all of $\lambda$ must vanish for the dilatino variation to be
zero.
To each of these we may add 8 unconstrained supercharges which are projected out by the
action of $\Gamma^+$.
This leads to the following counting of supercharges:
For type IIA, we have the standard 16, together with 6 from the left handed \susy\
parameter and zero from the right handed one, for a total of 22 supercharges.
In type IIB, if both spinors are left handed, we have the maximal of 28 supercharges
for a non-trivial \para\ background, while if
both are right handed, we only have 16, which is the minimum.
Clearly, for type IIA, the number of supercharges must be one of
$(16,18,20,22,24)$,
while for type IIB the allowed number of supercharges is $(16,20,24,28)$.
The sensitivity to chirality appears since the
non-zero components of right/left handed spinors appearing in
\eqref{right:chirality} and \eqref{left:chirality}
are required to satisfy different
sets of algebraic equations, whose form is evident in \eqref{H:contraction:explicit}.

Various configurations of the field strength, with their degrees of supersymmetry,
are presented in tables 1 and 2,
up to relabeling of the coordinates and parity.
In both tables,
unless otherwise stated, the $a_i$ are assumed non-zero and not to be equal.
We also state the symmetry of the transverse part of the given background, up to translations
which do not depends on the form of the constant field strength, and parity.
Since $16,20,24$ supercharges are possible in both IIA and IIB, there exists the
possibility of finding self T-dual backgrounds for these numbers of preserved supersymmetries.
This is discussed in section \ref{t:duality}.
The Penrose-Gueven limit of the $AdS_3 \times S^7$ geometry, presented in 
section 3.2,
has a background symmetry of
U(1) $\times$ U(3).
The maximal symmetry for a \ppp\ is U(4) and the minimal is U(1)$^4 \times \mathbb{Z}_3$.

\begin{table} \label{table:IIA}
\centering
\begin{tabular}{|c|c|c|} \hline\hline
$H_{+ij}$ components & Number of supercharges & Symmetry \\ \hline
$a_1=a_2=a_3=a_4=0$ (flat space) & 32 & O(8) \\ \hline
$a_1=\pm a_2,a_3=a_4=0$ & 24 & U(2) $\times$ O(4) \\ \hline
$a_1=\pm a_2=\pm a_3=\pm a_4$ & 22 & U(4) \\ \hline
$a_1=\pm a_2,a_3=\pm a_4$ & 20 & U(2)$^2 \times \mathbb{Z}_2 $ \\ \hline
$a_1=a_2+a_3,a_4=0$ & 20 & U(1)$^4 \times \mathbb{Z}_2$ \\ \hline
$a_1=\pm a_2 \pm a_3 \pm a_4$ & 18 & U(1)$^4 \times \mathbb{Z}_3$ \\ \hline
$a_1=a_2=a_3, a_4$ arbitrary & 16 & U(1) $\times$ U(3) \\ \hline
$a_1,a_2,a_3,a_4$ arbitrary & 16 & U(1)$^4 \times \mathbb{Z}_4$ \\ \hline
\hline
\end{tabular}
\caption{Classification of various \para\ backgrounds in non-chiral type IIA.}
\end{table}

\begin{table} \label{table:IIB}
\centering
\begin{tabular}{|c|c|c|} \hline\hline
$H_{+ij}$ components & Both left-handed & Both right-handed \\ \hline
$a_1=a_2=a_3=a_4=0$ (flat space) & 32 & 32 \\ \hline
$a_1=\pm a_2 \ne 0,a_3=a_4=0$ & 24 & 24 \\ \hline
$a_1=a_2=a_3=a_4 \ne 0$ & 28 & 16 \\ \hline
$-a_1=a_2=a_3=a_4 \ne 0$ & 16 & 28 \\ \hline
$a_1=-a_2=-a_3=a_4 \ne 0$ & 28 & 16 \\ \hline
$a_1=+a_2 \ne 0,a_3=-a_4 \ne 0$ & 16 & 24 \\ \hline
$a_1=a_2\ne 0,a_3=a_4\ne 0$ & 24 & 16 \\ \hline
$a_1=-a_2\ne 0,a_3=a_4\ne 0$ & 16 & 24 \\ \hline
$a_1=-a_2\ne 0,a_3=-a_4\ne 0$ & 24 & 16 \\ \hline
$a_1=a_2=a_3 \ne a_4$ & 16 & 16 \\ \hline
\hline
\end{tabular}
\caption{Classification of various \para\ backgrounds in chiral type IIB}
\label{diag}
\end{table}


Flipping the sign of one of the $a_i's$ corresponds to a parity operation on
the two dimensional plane whose field strength components $a_i$ determines,
since the field $H_{abc}$ is odd under parity. This parity acts by changing the sign
of one of the coordinates on the plane; changing both amounts to a rotation.
This parity operation also interchanges the chirality of the spinors,
and so the number of supercharges is insensitive to this sign in IIA.
In type IIB, this interchange corresponds to the exchange of two solutions in table
2, taking solutions with both left handed spinors to both right handed
and vice-versa. For a few exceptional cases with some $a_i$ vanishing,\footnote{
See footnote (\ref{foot:label}).}
the IIB theory of one chirality has the same amount of supersymmetry as the IIB
theory of opposite chirality,
and so the number of supercharges, even for type IIB, is not sensitive to parity.
An example is the background $a_1=\pm a_2,a_3=a_4=0$.

\subsection{Penrose limit of $AdS_{3}\times S^{7}$}
\label{ads3xs7}

Given any \sugra\ solution one can take the Penrose limit and obtain a 
plane-wave geometry. We may start with a \para\ \sugra\ solution. Then, it 
is easy to show that \party\ survives the Penrose 
limiting procedure, and hence after the Penrose limit we find a \ppp , 
which is necessarily of the form given in (\ref{pppwave2}, \ref{muij}). A 
well-known 
example is the Penrose limit of $AdS_{3}\times S^{3}\times \mathbb{R}^4$ 
\cite{BMN, Russo}, which leads to a \ppp\ with $a_1=a_2$, $a_3=a_4=0$ in 
the notation of eq. (\ref{foura}). 

As another example, we work out the Penrose limit of $AdS_{3}\times 
S^{7}$ manifold discussed earlier in section 3.2.
To begin with, for definiteness, let us consider the $AdS_{3}\times S^{7}$ 
case. To take the Penrose limit, it is more convenient to write the metric 
in the $AdS$ global coordinates:
\begin{subequations}
\label{Global}
\begin{align}
ds^2 &= R_1^2 [-\cosh^2\rho d\tau^2 +d\rho^2+\sinh^2\rho d\phi^2]+
R_2^2 [\cos^2\psi d\theta^2 +d\psi^2+\sin^2\psi d\Omega_5^2] \\
H_{\tau\rho\phi} &= 2 R_1^2\cosh\rho\sinh\rho
\end{align}
\end{subequations}
where
\be \label{define:r}
\frac{R_2^2}{R_1^2}\equiv r^2
\ee
and the other $T$-field components along
$S^7$ given in equation (\ref{S7:structure}).
Now we take the $R_1\to\infty$ limit together with
\bea\label{Penrose}
\rho=\frac{y}{R_1}\ &,&\ \ \ \psi=\frac{x}{R_2}\ , \cr
u=\tau+r\theta \ &,& \ \ \ 
v=\frac{1}{2R^2_1}(\tau-r\theta)\ ,
\eea
where $x,y,u,v$, all the other coordinates and also $r$ are kept fixed. 
Then the 
metric (\ref{Global}) becomes
\begin{subequations}
\label{afterlimit}
\begin{align}
ds^2  =  -2 du dv  - 
\left( \frac{1}{r^2}\sum_{i=1}^6{x_i}^2+\sum_{a=1}^2{y_a}^2 \right) du^2 \: &+ \:
\sum_{i=1}^6 \: dx^i dx^i+ \sum_{a=1}^2 \: dy^a dy^a \ , \\
H_{+12} = 2\ ,\ H_{+34} = H_{+56} &=H_{+78}=\frac{2}{r} \ .
\end{align}
\end{subequations}
This solution, in the notation of (\ref{foura}), is
$a_1=1,\ a_2=a_3=a_4=\frac{1}{r}$.
As discussed in section 3.2, $AdS_3\times S^7$, due to the 
fact that the parallelizing torsion on $S^7$ is not a closed form, 
is not a \sugra\ solution. However, one might still take its Penrose limit.
The novel point is that the non-closed parts of the 
torsion will drop out after the limit and hence the corresponding 
plane-wave is a \ppp\ solution of \sugra .

Using our previous arguments, one can show that the \ppp\ 
(\ref{afterlimit}) only preserves the 16 kinematical \susys .
As shown in \cite{Figueroa-O'Farrill:2002ft}, for a background
to preserve more than half the maximal supersymmetry, it must 
have a trivial dilaton dependence. These are generic behaviours of \sugra\ 
solutions.
The Penrose-Gueven limit can never destroy any supersymmetry, but
might enhance it \cite{Blau}.
 

One might wonder whether Ricci-\party\ is enhanced 
to \party\ under the Penrose limit. This is not true in 
general. As an example, if we start with $AdS_3\times S_q^7$ as a 
Ricci-\para\ geometry, it remains only Ricci-\para\ 
after the Penrose limit. 

\section{T-duality on parallelizable pp-waves}
\label{t:duality}

In this section we study T-duality on the \para\ \pp s  introduced in 
previous sections. In general, in order to perform T-duality, we first need to 
compactify the manifold and for (toroidal) compactifications of any 
manifold we require translations  along the compactification 
directions to be  (space-like) isometries of the manifold.
The existence of translation isometry along a space-like direction for the 
\ppp\, as written 
in eq. (\ref{pppwave2}), is not manifest.
However, the plane-wave solutions generically possess non-linearly 
realized symmetries and there is a chance that in a proper coordinate 
system some isometries which are hidden may become manifest. 
This is indeed the case. To see this, following Michelson 
\cite{Michelson}, let us consider the ``rotating'' frame
\bea
X^1=x^1 \cos (a_1 u) -x^2 \sin (a_1 u) \ , \cr
X^2=x^2 \cos (a_1 u) +x^1 \sin (a_1 u) \ ,
\eea
leaving all the other coordinates unchanged, i.e.
\be
U=u\ , \ \ {\tilde V}=v\ ,\ \ X^i=x^i\ \ i \geq 3\ .
\ee
The metric in the new coordiantes becomes
\bea\label{T1metric}
  ds^2 &= &-2  dU d {\tilde V} + dX^1dX^1+ dX^2 dX^2+ \sum_{i=3}^8 dX^i 
dX^i\\ 
&-&\!\!\! [a_2^2 (X_3^2 + X_4^2) +a_3^2 (X_5^2 + X_6^2) +a_4^2 (X_7^2 + 
X_8^2)] 
dU^2  
-2a_1 (X^1dX^2-X^2dX^1)dU \ ,
\nonumber
\eea
while the $H$-field remains invariant, i.e. $H_{UX^1X^2}=2a_1$.
It is evident that we can make the same transformation  
for all the other coordiantes and remove the $dU^2$ term of the metric 
completely; then we get a metric which has $dUdX^i$ term proportional to 
$F_{ij}X^j$. The translational symmetry is not manifest yet, to see that 
let us redefine\footnote{
We remind the reader that, as far as the metric is concerned, 
to make the isometry along a direction, say $X^1$, manifest, the 
frame we choose can be rotating clockwise or counter-clockwise. However, 
due to the presence of a $B$-field in our case, there is a preferred 
orientation.}
\[
V={\tilde V} - a_1 X^1 X^2 \ ,
\]
and the metric becomes
\bea\label{T2metric}
  ds^2 &=& -2  dU d{V}  + \sum_{i=1}^8 \: dX^i dX^i \cr
&& - [a_2^2 (X_3^2 + X_4^2) +a_3^2 (X_5^2 + X_6^2) +a_4^2 (X_7^2 + 
X_8^2)] dU^2  +4a_1 X^2dX^1dU \ .
\eea
Now we are ready to compactify $X^1$ on a circle and use Bucher's 
rules to perform T-duality \cite{Berg}
\bea
g^T_{\mu\nu}&=&g_{\mu\nu}-\frac{1}{g_{11}}\left(
{g_{\mu 1}g_{\nu 1}-B_{\mu 1}B_{\nu 1}}\right) \cr
B^T_{\mu\nu}&=&B_{\mu\nu}+2\frac{1}{g_{11}}\left({g_{\mu 1}B_{\nu 
1}-B_{\mu 1}g_{\nu 1}}\right) \cr
g^T_{\mu 1}&=&\frac{B_{\mu 1}}{g_{11}}\ ,\ 
B^T_{\mu 1}=\frac{g_{\mu 1}}{g_{11}}\ ,
\eea
In our case\footnote{ Given $H=dB$ there is a $U(1)$ gauge freedom in the 
definition of $B$ and we choose the gauge which is compatible with the 
translational symmetry along $X^1$.}
\be\label{TB}
B_{\mu 1}= g_{\mu 1}= 2a_1 \ X^2 \delta_{\mu U}\ .
\ee
Choosing $g_{11}=1$, i.e., sitting at the self-dual radius, we find
\bea
B^T_{\mu\nu}=B_{\mu\nu}\ &,& \ g^T_{\mu\nu}=g_{\mu\nu}, \cr
B^T_{\mu 1}=g_{\mu 1}\ &,& \ g^T_{\mu 1}=B_{\mu 1} .
\eea
As we see, the $g$ and $B$ fields are exactly the same before and after 
T-duality; in other words  \ppp s are invariant under T-duality.
 This invariance is a direct consequence of the specific form of our 
metric and $B$-field, namely, $B_{\mu 1}= g_{\mu 1}$, which is dictated by 
\party .

So far, we have shown that any \ppp\  solution of type IIB is also a 
solution of IIA, related by T-duality. In the above 
arguments we have only considered the bosonic fields $g$ and $B$.
One should also consider fermions and check if the above
T-duality invariance also holds in the fermionic sector. 
First, we note that the two 
conserved supersymmetries of the IIB 
background have the same chirality, while those of IIA have 
different chiralities. Compactification imposes a boundary condition 
on fermions, which is not necessarily compatible with their chirality and 
as a result, the number of \susys\ may change under T-duality.\footnote{
In other words one should check if eq.(\ref{ghat}) has the right 
periodicity condition.} While individual supersymmetries may be affected 
by compactification,
the total number of kinematical \susys\ (16) is not, and hence
16 such supercharges survive compactification and T-duality.
The difference between IIB and IIA only arises
in the non-kinematical \susys .  
If the number of supercharges of a type IIA
\ppp \ is $N_A=4k+2$ ($k=4,5$), then the corresponding T-dual 
IIB solution  
has $N_B=N_A\pm (N_A-16)$; the $+/ -$ depends on the ``orientation'' with
respect to the H-field.
Because of the existence 
of a non-trivial H-field, the two different orientations on  the 
compactification 
circle lead to two different solutions, differing by the number of 
\susys . For the $N_A=4k$ ($k=4,5,6$) cases, after T-duality we 
have $N_B=N_A$ or $N_B=N_A\pm (N_A-16)$. Which of these 
cases we obtain depends on the details of the \sugra\ solution. 
Therefore, invariance under 
T-duality can only be exact (in the sense that it holds in both 
bosonic and fermionic sectors) for $N_A=N_B=16$,  
$N_A=N_B=20$ and $N_A=N_B=24$. Note, however, that for $N_A=20, 24$ cases 
T-duality should be performed along the direction with $a_i=0$.
As examples of these  cases we
mention the \pl\ coming as the Penrose limit of $AdS_{3}\times S^{7}$ 
($N_A=N_B=16$) and $AdS_{3}\times S^{3}\times T^4$ ($N_A=N_B=24$) 
solutions.

\section{String theory on parallelizable pp-waves}

As noted in \cite{Metsaev}, the string theory (sigma model) on a generic 
homogeneous \pl\ is solvable in the light-cone gauge. In \cite{Metsaev}
the interest was mainly in backgrounds with
non-zero RR flux. In this section we formulate string theory on \ppp s
and show that, in this case, string theory is simpler than the 
generic plane-wave with RR flux. The formulation of string theory on some 
special \ppp s, namely those coming from the Penrose limit of $AdS_3\times 
S^3$ and its variants, have been previously considered in \cite{Russo, 
Ganor1, PhD}, although the connection to \party\ was not made.
We first focus on the bosonic sector of strings and in the next
section study fermions in the Green-Schwarz (GS) formulation.
However, since our backgrounds are only 
in the NS-NS sector, the RNS formulation can also be used. 

\subsection{Bosonic sector}

The non-linear sigma model in the NS-NS background of 
$G_{\mu\nu}$ and $B_{\mu\nu}$ fields is given by
\be
S=\frac{-1}{4\pi\alpha'}\int d^2\sigma\ \sqrt{-g} \left(
g^{ab} G_{\mu\nu} \partial_a X^\mu\partial_b X^\nu
+\epsilon^{ab} B_{\mu\nu} \partial_a X^\mu\partial_b X^\nu\right)\ ,
\ee
where $a,b=1,2$ and $g_{ab}$ is the worldsheet metric. For the background 
defined through eqs. (\ref{pppwave2},\ref{muij},\ref{foura}), we have
\be
\begin{split}
\label{sigmamodel}
S=\frac{-1}{4\pi\alpha'}\int d^2\sigma \sqrt{-g} \Bigl[
g^{ab} \bigl(-2\partial_a U\partial_b V
&-\frac{1}{4} h_{ik}h_{jk} X^i X^j \partial_a U\partial_b U+
 \partial_a X^i\partial_b X^i \bigr) \\
&+\epsilon^{ab} h_{ij} X^j \partial_a U\partial_b X^i\Bigr]
\ .
\end{split}
\ee

We next fix the conformal symmetry in the gauge
$g_{\tau\sigma}=0$ and $-g_{\tau\tau}= g_{\sigma\sigma}=1$, and to avoid 
ghosts we choose light-cone gauge, in which
\be\label{lightconegauge}
\partial_\tau U=p^+=const.
\ee
Then $V$ is constrained to satisfy
\bea\label{Vconstraints}
p^+\partial_\sigma V &=& \partial_\tau X^i \partial_\sigma X^i\ , \cr
2p^+\partial_\tau V &=& \partial_\tau X^i \partial_\tau X^i +
\partial_\sigma X^i \partial_\sigma X^i - \frac{1}{4}(p^+)^2 
h_{ik}h_{jk}X^iX^j\ .
\eea
Plugging the above into the action (\ref{sigmamodel}), the light-cone 
action for the transverse modes $X^i$ is seen to be
\bea\label{LCaction}
S_{LC} &=& \frac{1}{4\pi\alpha'}\int d^2\sigma \left[
\partial_\tau X^i\partial_\tau X^i -\partial_\sigma X^i\partial_\sigma X^i
-\frac{{p^+}^2}{4} h_{ik}h_{jk} X^i X^j 
- p^+ h_{ij} X^j \partial_\sigma X^i\right] \cr
 &=& \frac{1}{4\pi\alpha'}\int d^2\sigma \left[
(\partial_\tau X^i)^2-(\partial_\sigma 
X^i+\frac{p^+}{2}h_{ij}X^j)^2\right]\ .
\eea
Without loss of generality we can take the $h_{ij}$'s as in 
equation (\ref{foura}). Since the analysis for the different modes is quite 
similar, here we only focus on the $X^1,X^2$ components, whose \eom\  are
\be\label{eomX}
(\partial_\tau ^2-\partial_\sigma ^2)X^i +{(a_1p^+)}^2 X^i 
-{2a_1p^+}\epsilon_{ij}\partial_\sigma X^j =0\ ,\ \ \ i,j=1,2\ .
\ee
The above equation is solved by $x_i e^{i(\omega^{\pm}_n \tau-2n\sigma)}$ 
where
\be\label{B-freq}
\omega^{\pm}_n=2n \pm a_1p^+ \ ,
\ee
and the closed string boundary conditions
\be\label{closedBC}
X^i(\sigma)=X^i(\sigma+\pi) \ ,
\ee
fixes $n$ to be integer. Therefore, the most generic solution to 
equation (\ref{eomX}) can be written as
\begin{subequations}
\label{Rightmovers}
\begin{align}
X_R^1&=\frac{1}{2} \sum_{n\in \mathbb{Z}} 
\frac{1}{\omega^+_n} \alpha^R_n e^{i(\omega^+_n \tau-2n\sigma)}+
\frac{1}{\omega^-_n} \beta^R_n e^{i(\omega^-_n \tau-2n\sigma)} \ , \\
X_R^2&=\frac{1}{2} \sum_{n\in \mathbb{Z}} 
\frac{+i}{\omega^+_n} \alpha^R_n e^{i(\omega^+_n \tau-2n\sigma)}+
\frac{-i}{\omega^-_n} \beta^R_n e^{i(\omega^-_n \tau-2n\sigma)} \ ,
\end{align}
\end{subequations}
for the right-movers, and for the left-moves
\begin{subequations}
\label{Leftmovers}
\begin{align}
X_L^1&=\frac{1}{2} \sum_{n\in \mathbb{Z}} 
\frac{1}{\omega^+_n} \alpha^L_n e^{i(\omega^+_n \tau+2n\sigma)}+
\frac{1}{\omega^-_n} \beta^L_n e^{i(\omega^-_n \tau+2n\sigma)} \ , \\
X_L^2&=\frac{1}{2} \sum_{n\in \mathbb{Z}} 
\frac{i}{\omega^+_n} \alpha^L_n e^{i(\omega^+_n \tau+2n\sigma)}+
\frac{-i}{\omega^-_n} \beta^L_n e^{i(\omega^-_n \tau+2n\sigma)} \ .
\end{align}
\end{subequations}
The reality of $X$'s implies that
\be\label{reality}
(\alpha_n^R)^\dagger=\beta_{-n}^R\ ,\ \ (\alpha_n^L)^\dagger=\beta_{-n}^L\ ,
\ee
and $X^i=X^i_R+X^i_L$.

Note that for a generic $a_1 p^+$, there is no zero-mode and the mode 
expansions of (\ref{Rightmovers},\ref{Leftmovers}) can be used without 
ambiguity. For $\frac{1}{2}a_1 p^+\in \mathbb{Z}$, however, we have zero 
frequency modes; for such cases one may extract the zero-modes:
\begin{subequations}
\label{integerp+}  
\begin{align}
X^1&=(x^1+p^1\tau)\cos a_1 p^+ \sigma+ (x^2+p^2\tau)\sin a_1 p^+\sigma + 
\sum_{n\neq
a_1 p^+/2} \ Oscil. \\ 
X^2&=(x^2+p^2\tau)\cos a_1 p^+\sigma - (x^1+p^1\tau)\sin a_1 p^+\sigma + 
\sum_{n\neq a_1 p^+/2} \ Oscil. 
\end{align}
\end{subequations}
In these cases, for a constant $\tau=\tau_0$ slice in the 
$X^1,X^2$ plane, the string forms a circle of radius 
$X_1^2+X_2^2=(x^1+p^1\tau_0)^2+(x^2+p^2\tau_0)^2$, with some wiggles superposed
on it.
To understand the physics of such strings it is helpful to work out the 
angular momentum of the center of mass of the string:
\be\label{angular}
L=\frac{1}{\pi}\int_0^\pi (X^1\partial_\tau X^2- X^2\partial_\tau 
X^1) d\sigma =x^1p^2-x^2p^1\ ,
\ee
which is constant. In this case strings are circles whose
radius grows with time, while carrying constant angular momentum.
This is as expected if we note that the NS-NS field acts like a magnetic field. 
However, the effect of $du^2$ terms in the background metric
appears in the fact that the $p^i\tau$ terms in (\ref{integerp+})
are multiplied by $\sin a_1p^+\sigma$ and $\cos a_1p^+\sigma$ factors, i.e.,
the center of mass of a string on the average has zero momentum and is 
confined to stay around $X=0$. Similar behaviour have been observed for 
other \pl s (e.g. see \cite{Metsaev}). 

One of the remarkable differences between our case and the other plane-waves
(which involve RR fluxes), is that the spacing between the energy 
levels in our case is just given by integers (in other words 
$\omega^{\pm}_n$ is a linear function of $n$).

We would like to point out that due to the presence of background 
fields, the light-cone Hamiltonian differs from $\int \partial_\tau V$, 
explicitly
\bea\label{LChamiltonian}
H_{LC}= \Pi_+ &=& 
\int d\sigma\ (\partial_\tau V + p^+a_1^2 X_i^2+ a_1\epsilon_{ij}X^i\partial_\sigma X^j)\ \cr
 &=&\frac{1}{2p^+} \int d\sigma\ \Bigl[(\partial_\tau X^i)^2 
+(\partial_\sigma X^i+p^+ a_1\epsilon_{ij}X^j)^2\Bigr]\ \cr
&=& \frac{1}{2p^+} \sum_{n\in \mathbb{Z}} 
\alpha_n^R\beta_{-n}^R+\beta_n^R\alpha_{-n}^R+ 
\alpha_n^L\beta_{-n}^L+\beta_n^L\alpha_{-n}^L\ .
\eea
Note that the last line of the above equation is written for 
$\frac{1}{2}a_1 p^+ \notin \mathbb{Z}$. For integer values of 
$\frac{1}{2}a_1 p^+$, one should add $(p^1)^2+(p^2)^2$ to the sum over 
oscillators.

Given the mode expansions, we can proceed with the quantization of strings by imposing
\bea\label{quant}
[X^i(\sigma),X^j(\sigma')]=0\ ,\ \ [P^i(\sigma),P^j(\sigma')]=0\ ,\ \ 
[X^i(\sigma),P^j(\sigma')]=i\delta^{ij}\delta(\sigma-\sigma')\ ,
\eea
with $P^i=\partial_\tau X^i$, leading to
\bea\label{modequant}
[\alpha^R_n,\beta^R_m]= [\alpha^L_n,\beta^L_m]= \omega_n^-\delta_{m+n} \ ,
\eea
and all the other combinations commuting. For the cases where we have ``zero-modes'' 
($\frac{1}{2}p^+a_1\in\mathbb{Z}$), $[x^i,p^j]=i\delta^{ij}$. The above 
mode expansion and commutators is similar to the twisted sector of strings on 
orbifolds.
Using the commutation relations (\ref{modequant}) we can write the
Hamiltonian in a normal ordered form
\be
H_{LC}= \frac{1}{2p^+} \sum_{i=1}^4\sum_{n \geq 0} 
\alpha_{-n\ i}^R\beta_{n\ i}^R+\beta_{-n\ i}^R\alpha_{n\ i}^R+ 
\alpha_{-n\ i}^L\beta_{n\ i}^L+\beta_{-n\ i}^L\alpha_{n\ i}^L+ E_0\ ,
\ee
where $E_0$ is the zero point energy coming from regularizing sums 
like $\sum (n-\phi)$, and has value
\be\label{E0}
E_0=-\frac{2}{3}-\sum_{i=1}^4 \phi^2_{i} \ ,
\ee
where $\phi_i$ are the non-integer part of $\frac{1}{2}p^+a_i$, i.e. 
$\phi_i=\frac{1}{2}p^+a_i-[\frac{1}{2}p^+a_i]$.

\subsection{Fermionic sector}
\label{fermionic:sector}

The fermionic NS-NS sector of the Green-Schwarz action, expanded to second order\footnote{
Terms of higher order do not contribute
in light-cone gauge.}
in the fermions, for type IIA string theory is
\be
\label{GSW:IIA}
S_F \: = \:\frac{i}{\pi \alpha^\prime}  \int d^2\sigma
\bar\theta  \beta^{ab} \partial_a X^\mu \Gamma_\mu \hat{D}_b \theta \ ,
\ee
with $\beta^{ab}=\sqrt{-h} h^{ab} \sigma_0 - \epsilon^{ab} \sigma_3$,
$\hat{D}_b$ the pull-back of the superspace covariant derivative
with torsion
\be
\hat{D}_a \: = \: D_a \: + \: \frac{1}{8} \partial_a X^\mu \Gamma^{\rho \sigma}
\sigma_3 H_{\mu \rho \sigma} \ ,
\ee
and the normal covariant derivative 
\be
D_a \: = \: \partial_a \: + \: \frac{1}{4} (\partial_a X^\mu) \omega_\mu^{ab}
\Gamma_{ab}
\ee
contains the spin connection $\omega_\mu^{ab}$.
The NS-NS sector of the IIB Green-Schwarz action is the same as that presented
above, with the difference in the solutions arising from the chirality of the
two spinors, as in the discussion of supersymmetry counting in section \ref{susy:counting}.

We work in light-cone gauge, where we impose the conditions
\be
X^+=p^+ \tau \ , \ \ \ \ \ \ \ \ \ \ \ \sqrt{-h} h^{ab}=\eta^{ab} \ ,
\ee
and fix the $\kappa$ symmetry by imposing the additional constraint
$\Gamma^+ \theta^\alpha=0$, giving
\be
S_F \: = \:\frac{- i p^+}{\pi \alpha^\prime}  \int d^2\sigma
\bar\theta  \Gamma^-
\left( \hat{D}_\tau \sigma_0 \: + \: \hat{D}_\sigma \sigma_3 \right)
\theta \ ,
\ee
with
\begin{subequations}
\begin{align}
\hat{D}_\sigma \: &= \:
\partial_\sigma \ , \\
\hat{D}_\tau \: &= \:
\partial_\tau \: + \: \frac{p^+}{4}
\left(
\omega_{+ \mu \nu} \: + \: \frac{1}{2} H_{+ \mu \nu}
\right) \Gamma^{\mu \nu} \ ,
\end{align}
\end{subequations}
yielding the following form for the fermionic action
\be
S_F \: = \:\frac{- i p^+}{\pi \alpha^\prime}  \int d^2\sigma
\sqrt{2} \left( \bar\theta^1 \Gamma^- \partial_+ \theta^1 +
\bar\theta^2 \Gamma^- \partial_- \theta^- \right)
-\frac{p^+}{8} \left(
\bar\theta^1 \Gamma^- \Gamma^{IJ} \theta^1
- \bar\theta^2 \Gamma^- \Gamma^{IJ} \theta^2 \right)
H_{+IJ} \ ,
\ee
where $I,J$ range over the transverse directions, and
$\partial_\pm=\frac{1}{\sqrt{2}}(\partial_\tau \pm \partial_\sigma)$.
Precisely half the components of the fermions
have been projected out by the gauge fixing to leave
only the physical degrees of freedom.
The equations of motion are
\begin{subequations}
\label{fermion:eom}
\begin{align}
\left( \partial_+ - \frac{p^+}{8\sqrt{2}} \Gamma^{IJ} H_{+IJ} \right) \theta^1 \: = \: 0 , \\
\left( \partial_- + \frac{p^+}{8\sqrt{2}} \Gamma^{IJ} H_{+IJ} \right) \theta^2 \: = \: 0 ,
\end{align}
\end{subequations}
which do not couple $\theta^1$ to $\theta^2$, but do mix components
within the individual fermions.
The closed string boundary conditions are
\be
\label{bc}
\theta^1,\theta^2 |_{\sigma=0} \: = \: 
\theta^1,\theta^2 |_{\sigma=\pi}.
\ee
We solve the equations of motion \eqref{fermion:eom} by
separating the solution in light-cone coordinates, then expanding the
solution into a complete set of functions. The ansatz is
\[
\theta^1 (\tau, \sigma ) = 
e^{i n_1 (\tau + \sigma)}\ e^{i m_1 (\tau - \sigma)}
\ \psi^1_{m_1,n_1}\ ,\  
\theta^2 (\tau, \sigma ) = 
e^{i n_2 (\tau - \sigma)}\ e^{i m_2 (\tau + \sigma)}
\ \psi^2_{m_2,n_2}\ ,\  
\]
with $\psi^1$ and $\psi^2$ constant spinors which are
subject to the gauge fixing condition.
Introducing the mass matrix $M=\frac{-i p^+}{16} \hplus$,
with the matrix
$\hplus$ written out in equation \eqref{H:contraction},
we have
\begin{subequations}
\label{eigenvalue:equations}
\begin{align}
\left( n_1 \: - \: M \: \right) \psi^1_{m_1,n_1} \: = \: 0 \ , \\
\left( n_2 \: + \: M \: \right) \psi^2_{m_2,n_2} \: = \: 0 \ ,
\end{align}
\end{subequations}
which are a pair of eigenvalue equations for $n_1$ and $n_2$.
Imposing the boundary condition \eqref{bc}, we see that
\be
m_1 \: \in \: 2 \mathbb{Z}+n_1 \ , \ \ \ \ \ \ \ \ \ \
m_2 \: \in \: 2 \mathbb{Z}+n_2 \ .
\ee
The final mode expansions are
\begin{subequations}
\begin{align}
\theta^1 (\tau, \sigma ) \: &=\:
\sum_{m_1} 
e^{i n_1 (\tau + \sigma)}\ e^{i m_1 (\tau - \sigma)}\ \psi^1_{m_1,n_1} \ , \\
\theta^2 (\tau, \sigma ) \: &=\:
\sum_{m_2} 
e^{i n_2 (\tau - \sigma)}\ e^{i m_2 (\tau +\sigma)}\ \psi^2_{m_2,n_2} \ . 
\end{align}
\end{subequations}
The reality condition implies that
\be
\left( \psi^\alpha_{m_\alpha,n_\alpha} \right)^\dagger \: = \:
\psi^\alpha_{-m_\alpha,-n_\alpha} \ .
\ee
Upon quantization 
\[
\{(\psi^\alpha_{m_\alpha,n_\alpha})^\dagger,\ 
\psi^\beta_{m'_\beta,n'_\beta}\}=\delta_{\alpha \beta}\delta_{mm'}\delta_{nn'} \ .
\]
It is straightforward to work out the light-cone Hamiltonian for the 
fermionic modes
\be
H_{LC}=\frac{1}{2p^+}\sum_{m_\alpha \geq 0} (\psi^\alpha)^\dagger m_\alpha \psi^\alpha- E_0 \ ,
\ee
where $E_0$ is the zero point energy and is defined in eq. (\ref{E0}). As 
we expect, the bosonic and fermionic zero point energies are the same 
up to a sign and hence there is no total zero point energy in the 
spectrum, a confirmation of the \susy\ of the background.

The \para\ solutions we are considering are parameterized by four
constants, as in \eqref{foura}.
As an example, consider the background with $-a_1=a_2=a_3=a_4$. This background
preserves 22 supercharges in type IIA and either 16 or 28 supercharges in type IIB,
depending on whether the spinors are chosen to be both left-handed or right-handed,
respectively.
The mass matrix for this configuration is
\be
M=\frac{a_1 \: p^+}{4} \: diag(1,1,-1,-1,-1,1,-1,1,0,2,-2,0,0,0,0,0) ,
\ee
where we only need to consider the half of the mass matrix acting on the subspace
surviving the gauge fixing.
Now, for the modes for which the mass vanishes, equation \eqref{eigenvalue:equations}
implies the existence of zero-modes.
After fixing the $\kappa$ symmetry by imposing the gauge $\Gamma^+ \theta^\alpha = 0$,
the counting of the zero-modes for the physical degrees of freedom
parallels the counting of the non-kinematical supercharges presented in section
\ref{susy:counting}. The number of zero-modes depends on whether we are working
with type IIA or IIB, and in type type IIB, whether we choose both spinors
to be left-handed or right-handed.  The number of zero-modes can be ascertained
by looking at tables 1 and 2, subtracting 16 for the standard kinematical supersymmetries,
which in the present discussion correspond to the modes orthogonal to those
removed by the gauge fixing. This leads to a subtlety, which is that the number of
zero-modes coming from left handed spinors $\theta$ here are associated with the number
of non-kinematical supercharges (for counting purposes only)
for right handed spinors in section \ref{susy:counting},
and vice-verse. The total counting in type IIA, of course, is the same.
Therefore, for this example, there are 6 zero-modes
in type IIA and zero (both spinors right-handed) or 12 (both left-handed)
zero-modes in type IIB.

The example $-a_1=1,a_2=a_3=a_4=1/r$,
with $r$ defined in \eqref{define:r}, arising as
the Penrose limit of $AdS_3 \times S^7$,
is half supersymmetric, preserving
16 supercharges in both type IIA and IIB, and as a result, the fermionic solutions
exhibit no zero-modes. The mass matrix is
\be
M=\frac{p^+}{8 r^2} \: diag(r^2-1,r^2-1,r^2+3,r^2-1,r^2+1,r^2-3,r^2+1,r^2+1) .
\ee

This analysis can be extended to other configurations, with the general result that the
number of zero-modes, in the solutions to the fermionic sector, is equal to the number of non-kinematical supercharges preserved by the background, taking account of the type IIB reversal
discussed above.

\section{Compactification and T-duality}

In section 5 we studied T-duality of \ppp s at the \sugra\ level. In this 
section we extend the T-duality analysis to string theory.
As we will see, because of the nature of the background, only   
right-movers contribute to the center of mass modes. This, and other peculiar features of 
strings under T-duality will be examined in this section.

We begin by writing the sigma model action in the \ppp\ after a Michelson 
rotation, i.e., the background given in eqs. (\ref{T2metric}),(\ref{TB}). 
After fixing the light-cone gauge:
\bea\label{TLCaction}
{\tilde S}_{LC} = \frac{1}{4\pi\alpha'}\int d^2\sigma \left[
\partial_\tau \tX^i\partial_\tau \tX^i -\partial_\sigma \tX^i\partial_\sigma \tX^i
-4 p^+ a_1 \tX^2 (\partial_\tau \tX^1+\partial_\sigma \tX^1)\right] \ ,
\eea
where we have only presented the action for the $\tX^1, \tX^2$ components, the 
other $X$'s are similar and here we skip them. The \eom\ are
\be\label{TeomX}
(\partial_\tau ^2-\partial_\sigma ^2)\tX^i 
-{2a_1p^+}\epsilon_{ij}(\partial_\tau \tX^j+\partial_\sigma \tX^j) =0\ ,\ \ \ 
i,j=1,2\ .
\ee
Inserting $x_i e^{i(\omega \tau-2n\sigma)}$
we find
\be\label{Tfreq}
\omega=2n \ \ \ {\rm or}\ \ \ \tomega^{\pm}_n=2n\pm 2a_1p^+\ .
\ee
Imposing the closed string boundary conditions restricts $n$ to the 
integers. Then the mode expansions are
\begin{subequations}
\label{TRightmovers}
\begin{align}
\tX_R^1&=\tx^1+\tp^1(\tau-\sigma)+\sum_{n\in \mathbb{Z}}
\frac{i}{2n} \talpha^R_n e^{2in(\tau-\sigma)} \ , \\
\tX_R^2&=\tx^2+\tp^2(\tau-\sigma)+\sum_{n\in \mathbb{Z}}
\frac{i}{2n} \tbeta^R_n e^{2in(\tau-\sigma)} \ ,
\end{align}
\end{subequations}
for the right-movers and\footnote{For $p^+a_1\in \mathbb{Z}$, there is a 
``zero-mode'' for left-movers, which should be extracted in a manner similar to
eq. (\ref{integerp+}). 
}
\begin{subequations}
\label{TLeftmovers}
\begin{align}
\tX_L^1&=\frac{1}{2} \sum_{n\in \mathbb{Z}}
\frac{1}{\tomega^+_n} \talpha^L_n e^{i(\tomega^+_n \tau+2n\sigma)}+
\frac{1}{\tomega^-_n} \tbeta^L_n e^{i(\tomega^-_n \tau+2n\sigma)} \ , \\
\tX_L^2&=\frac{1}{2} \sum_{n\in \mathbb{Z}}
\frac{i}{\tomega^+_n} \talpha^L_n e^{i(\tomega^+_n \tau+2n\sigma)}+
\frac{-i}{\tomega^-_n} \tbeta^L_n e^{i(\tomega^-_n \tau+2n\sigma)} \ .
\end{align}
\end{subequations}
for the left-movers.
We see that only the right movers have the usual momentum mode. Note also 
that for the non-compact case, 
$\tp^1=\tp^2=0$, resulting from the boundary conditions, 

Next, we quantize the strings by imposing the usual commutation relations
(\ref{quant}), where now the conjugate momenta to $\tX^i$ are
\begin{subequations}
\label{Tmomenta}
\begin{align}
\tP^1&= \partial_\tau\tX^1-2p^+a_1 \tX^2 \ , \\
\tP^2&= \partial_\tau\tX^2 \ .
\end{align}
\end{subequations}
The canonical quantization conditions lead to
\bea\label{Tmodequant}
[\tx^1,\tx^2]=-\frac{i}{2p^+a_1}\ \ \ &,& [\tx^i ,\tp^j]=0\ ,\ \ \ [\tp^1,\tp^2]=0\ , \cr
[\talpha^R_n,\talpha^R_m]=[\tbeta^R_n,\tbeta^R_m]=\frac{-4n^3}{4n^2-({p^+}a_1)^2}\delta_{m+n}\ 
&,&\ \ 
[\talpha^R_n,\tbeta^R_m]=ip^+a_1\frac{2n^2}{4n^2-({p^+}a_1)^2}\delta_{m+n}\ , \cr
[\talpha^L_n,\talpha^L_m]=[\tbeta^L_n,\tbeta^L_m]=0\ & , & \ \ 
[\talpha^L_n,\tbeta^L_m]=\frac{4(n+p^+a_1)^2}{2n+{p^+a_1}}\delta_{m+n}\ .
\eea
We note that the ``zero-modes'', $\tx^i,\tp^i$, have unusual commutation relations, e.g.
$[\tx^i,\tp^j]=0$, and also the zero-modes of $\tX^1$ and $\tX^2$ are 
non-commuting.
  
We would like to stress that in the rotating frame coordinates, the 
light-cone constraints are different from the usual ones
\begin{subequations}
\label{TVconstraints}
\begin{align}
p^+\partial_\sigma \tilde{V} &= \partial_\tau \tX^i \partial_\sigma \tX^i\ 
+2p^+a_1 \tX^2\partial_\sigma \tX^1 \ , \\ 
2p^+\partial_\tau \tilde{V} &= \partial_\tau \tX^i \partial_\tau \tX^i +
\partial_\sigma \tX^i \partial_\sigma \tX^i +4p^+a_1 \tX^2\partial_\tau \tX^1\ , 
\end{align}
\end{subequations}
and the light-cone Hamiltonian is
\be
\tilde{H}_{LC}=\frac{1}{2p^+}\int d\sigma \left[(\partial_\tau \tX^i)^2+
(\partial_\sigma \tX^i)^2 +4p^+a_1 \tX^2\partial_\sigma \tX^1\right] .
\ee

Now we proceed with compactification. Although from the solutions \eqref{TRightmovers}
and \eqref{TLeftmovers},
it may seem that both the $\tX^1$ and $\tX^2$ directions can have zero-modes (i.e., we have 
translational symmetry along $\tX^1$ and $\tX^2$ at the level of the \eom \eqref{TeomX}), 
it is only possible to compactify the system along the $\tX^1$ direction. This can be seen from
equations  \eqref{Tmomenta} or \eqref{TVconstraints}.
Putting $\tX^1$ on a circle of radius $R_1$, $\tp^1$ can become non-zero as a winding mode
\be
\tp^1=w_1 R_1\ ,\ \ \ w_1\in \mathbb{Z}\ ,
\ee 
while $\tp^2$ is still zero as a result of the closed strings boundary conditions.

Since $[\tx^1,\tp^1]=0$, there is no extra quantization condition on $\tp^1$ due to the 
momentum modes. In fact the momentum conjugate to $\tx^1$ is $\tx^2$ and it should 
have a discrete spectrum.
Noting that $\tx^1,\tx^2$ are the center of mass coordinates of strings along $\tX^1, 
\tX^2$, $[\tx^1,\tx^2]\neq 0$ means that the $\tx^1,\tx^2$ space is a non-commutative cylinder 
where $\tx^2$ is along the axis \cite{Chaichian}, its radius is $R_1$ and the fuzziness is
proportional to $\frac{1}{2p^+a_1}$. As discussed in \cite{Chaichian}, $\tx^2$ has a
discrete spectrum, $\tx^2\sim \frac{m_1}{2R_1p^+a_1}$ with $m_1\in \mathbb{Z}$. Upon 
T-dualizing along $X^1$ and exchanging $R_1$ with $\frac{1}{R_1}$, it is easy to show that 
the Hamiltonian remains unchanged if we also exchange $m_1$ with the winding $w_1$. 
In the fermionic sector as usual, duality acts by changing the chirality of one of the
fermions. As we have argued in previous sections, this may change the number of 
supersymmetries.

\section{D-Branes on parallelizable plane-waves}

In this section we study the existence of BPS $Dp$-branes in \ppp s
backgrounds. $Dp$-branes have been studied in various plane-wave 
backgrounds (e.g. see \cite{DP, Green, Michishita, Ganor1, KimLeeYang}). In 
general 
there are two ways of addressing the question of $D$-branes, one is 
through open strings with Dirichlet boundary conditions, as it was first 
introduced in \cite{Polch}, or through closed strings and the boundary 
state formulation
\cite{Callan}.  Here we follow the construction via open strings.  As we 
will show, due to presence of the $B$-field, three different situations can arise.

To start with, let us focus on the bosonic modes. First we note that, as
is evident from eq. (\ref{lightconegauge}),  
the light-cone gauge condition fixes the boundary condition the or $U$  
component to be Neumann. Therefore, $U$ lies inside the worldvolume of all 
$Dp$-branes we are going to study. As for the other components, we 
start with the light-cone action (\ref{LCaction}). This leads to the \eom\ 
eq. (\ref{eomX}) and the boundary conditions
\be\label{BC}
\int d\tau \delta X^i (\partial_\sigma X^i +p^+ a_1\epsilon^{ij} 
X_j)|_{\sigma=0}^{\pi}\ .
\ee
Since the argument for $X$'s along the directions where there is an 
$H$-field present is similar, here we consider $H_{+12}$ and the $(1,2)$ 
plane only. If some of the $a_i$'s in (\ref{foura}) are zero, the 
situation is of course the same as for flat space.

One can recognize three possibilities:
\begin{itemize}
\item[{\it i})] $X^1$ and $X^2$ are both transverse to brane.
\item[{\it ii})] $X^1$ and $X^2$ are both inside the brane.
\item[{\it iii})] Only $X^1$ or $X^2$ is along the brane brane.
\end{itemize}

Since these cases have been mentioned in \cite{Michishita}, we will be 
very brief with them.

{\it i}) This case is realized by imposing Dirichlet boundary conditions 
on both $X^1$ and $X^2$, i.e., $\delta X^i|_{\sigma=0,\pi}=0$. The boundary 
conditions force the frequency of the string modes to be integer valued 
and hence
\begin{subequations}
\begin{align}
X^1\!\!&=\!\! (x_1\cos p^+a_1\sigma- x_2\sin p^+a_1\sigma)\sigma 
+\sum_{n\neq 0}
\frac{\sin n\sigma}{n} \left( 
\alpha_n \cos (n\tau+p^+a_1\sigma)-\beta_n \sin 
(n\tau+p^+a_1\sigma)\right) \nonumber \\
X^2\!\!&=\!\! (x_2\cos p^+a_1\sigma+ x_1\sin p^+a_1\sigma)\sigma 
+\sum_{n\neq 0}
\frac{\sin n\sigma}{n} \left( 
\alpha_n \sin (n\tau+p^+a_1\sigma)-\beta_n \cos 
(n\tau+p^+a_1\sigma)\right) \nonumber
\end{align}
\end{subequations}
\addtocounter{equation}{-1} 
(Once $p^+a_1\in \mathbb{Z}$ we have extra zero-modes.)
Upon quantization we find
\be
[\alpha_n,\alpha_m]=[\beta_m,\beta_n]=n\delta_{m+n}\ .
\ee
In this case the $V$ direction also satisfies a Neumann boundary condition 
and hence lies inside the brane. The brane is located in such a way that $H_{+12}$ 
only has one leg along the brane. As discussed in \cite{Ganor1,Ganor2}, 
the theory living on such branes is a ``dipole'' gauge theory. As seen 
from the above mode expansions, these branes are stuck at $X^1=X^2=0$. 
In general, following \cite{DP, Michishita}, it is straightforward 
to show that these branes preserve half of the kinematical and half of the
non-kinematical supersymmetries.

{\it ii)} In this case, to satisfy the boundary conditions we demand 
that
\be\label{NBC}
\partial_\sigma X^i +p^+ a_1\epsilon^{ij} 
X_j|_{\sigma=0, \pi}=0\ ,
\ee
which is a modified Neumann boundary condition. It is  
well known that the existence of a background $B$-field will change the boundary 
conditions. The $V$ direction should also 
satisfy a modified boundary condition
\be\label{VNBC}
\partial_\sigma V +p^+ a_1\epsilon^{ij} 
X_j\partial_{\tau}X^j|_{\sigma=0, \pi}=0\ .
\ee
The modified boundary conditions (\ref{NBC}) and (\ref{VNBC}) may be 
understood from the fact that the brane now contains $U,V,X^1,X^2$, and the 
$H_{+12}$ field (which lies completely inside the brane) can be treated as a
background electric field equal to $p^+a_1 x^2$ on the brane. For the 
low energy theory on such branes we expect to find just a gauge theory with 
a non-constant background electric field. 

The frequencies of the normal modes satisfy the boundary conditions 
(\ref{NBC}), similar to the case {\it i)}, and are found to be integer valued 
and hence the mode expansions are
\begin{subequations}
\begin{align}
X^1&=x_1\cos p^+a_1\sigma- x_2\sin p^+a_1\sigma
+\sum_{n\neq 0}
\frac{ \cos n\sigma}{n} \left( 
\alpha_n \cos (n\tau+p^+a_1\sigma)-\beta_n \sin 
(n\tau+p^+a_1\sigma)\right) \nonumber \\
X^2&=x_2\cos p^+a_1\sigma+ x_1\sin p^+a_1\sigma
+\sum_{n\neq 0}\frac{\cos n\sigma}{n}  \left( 
\alpha_n \sin (n\tau+p^+a_1\sigma)-\beta_n \cos 
(n\tau+p^+a_1\sigma)\right) \nonumber
\end{align}
\end{subequations}
\addtocounter{equation}{-1} 

{\it iii)} In this case, we demand that $X^1$ satisfy Neumann b.c.'s, 
$\partial_{\sigma}X^1|_{\sigma=0,\pi}=0$ and $X^2$ to satisfy Dirichelet b.c.'s,
$\delta X^2=0$. It is not hard to check that, as a result, $V$ 
must satisfy the usual Neumann boundary condition, 
$\partial_{\sigma}V|_{\sigma=0,\pi}=0$.
One of the differences between this case and the previous two cases is 
that the frequencies of the string modes are now non-integer  
\be\label{openstringfreq}
\omega^{\pm}_n=n\pm p^+a_1 \ ,
\ee
with the mode expansions
\begin{subequations}
\label{openstrings}
\begin{align}
X^1&=\frac{1}{2}\sum_{n\in \mathbb{Z}}\left( 
\frac{1}{\omega^+_n} \alpha_n e^{i\omega^+_n \tau}-
\frac{1}{\omega^-_n} \beta_n e^{i\omega^-_n \tau}\right)\cos 
n\sigma \ , \\
X^2&=\frac{1}{2} \sum_{n\in \mathbb{Z}} \left(
\frac{1}{\omega^+_n} \alpha_n e^{i\omega^+_n \tau}
+\frac{1}{\omega^-_n} \beta_n e^{i\omega^-_n \tau}\right)\sin 
n\sigma \ , 
\end{align}
\end{subequations}
with the reality condition $\beta_n^\dagger=\alpha_{-n}$.
Imposing the quantization conditions (\ref{quant})
leads to
\be\label{openquant}
[\alpha_n,\beta_m]=\omega_n^+ \delta_{m+n}\ .
\ee
We should note that the above mode expansion is for $p^+ a_1\notin 
\mathbb{Z}$. For integer values of $p^+a_1$, however, we have a zero-mode
and the expansion is 
\begin{subequations}
\label{openintegerp+}
\begin{align}
X^1&= (x+p\tau)\cos p^+a_1\sigma+ \sum_{n\neq p^+a_1} Oscil. \\
X^2&= (x+p\tau)\sin p^+a_1\sigma+ \sum_{n\neq p^+a_1} Oscil.
\end{align}
\end{subequations}
where upon quantization we obtain $[x,p]=i$.

In this case the brane is located so that the $H_{+12}$ field has two legs 
along the brane and one transverse to it. In the gauge 
$B_{+1}=p^+a_1 X^2$, the $B$-field resides completely inside the brane, 
however, the value of the $B$-field is zero exactly on the brane which is 
necessarily at $X^2=0$. Again the $B_{+1}$-field can be understood as a 
background electric field on the brane, which is now proportional to one 
of the scalar fields (the ``transverse'' directions to the brane). This in 
particular gives a mass to that scalar field, so that its lowest 
excitation is no longer massless. This can also be observed from the 
mode expansion (\ref{openstrings}). A more detailed analysis of the 
theory living on these branes is postponed to future works.    

In principle, one can also work out the fermionic modes, but since the 
computations are very similar to those which appear in \cite{Michishita}, we do 
not present them here.

Finally, we would like to note that it is possible to have a combination of 
the three cases discussed above. For example, in the the background $a_1=a_2=a_3=a_4\neq 0$,
with 28 supersymmetries, we might have a D5-brane along 
the $UV1345$ directions, a mixture of all three cases.

\section{Discussion}

In this paper we classified and studied, to some extent, the \ppp s.
We first briefly studied 
implications of \party\ for a general \sugra\ solution and proved that the 
vanishing of the gravitino variation for \para\ backgrounds can be solved with 32
independent solution. One should note that the converse is not true,
that is, not all the cases for which the gravitino variation has 32 
solutions are parallelizable, for example, the famous $AdS_5\times S^5$ 
background.
 
 
Strings on $AdS_3$ have been studied in detail using the SL(2,$\mathbb{R}$)
group manifold and the corresponding Kac-Moodi algebra \cite{MO}. 
It would be very interesting to extend the definition of the 
WZW models and the Maldacena-Ooguri setup to the $S^7$ case; although it 
is not a group manifold, it has a nice (non-associative) algebraic (octonionic)
structure.

We then turned to \ppp s and proved that all \para\ \pp s are 
necessarily homogeneous \pl s and the converse is true if the $\mu_{ij}$ 
in the $du^2$ term ($\mu_{ij}x^ix^j$) has doubly degenerate eigenvalues.
The \ppp s may be classified by their \susy , where for type IIB
the maximal \susy\ is 28 and the others differ by steps of four (down to 16), 
while type IIA \susy\ 
may have 24, 22, 20, 18 and 16 supercharges. 
We also discussed the invariance of the bosonic sector of \ppp s under 
T-duality and discussed how the fermions and number of supercharges may change 
under T-duality.

We studied string theory on the \ppp\ backgrounds and showed that the
sigma model is simpler than for other \pl\ backgrounds.
This simplicity might help in working out the vertex operators, making it possible
to study string scattering processes and to evaluate the S-matrix elements, whose
existence for \pl \ backgrounds has been argued in \cite{Smatrix}.
Working out the proper vertex operators and string scattering amplitudes
is another interesting open question we postpone to future works. 

We also very briefly discussed the half BPS $Dp$-branes and the 
restrictions on the possible Dirichelet or Neumann boundary conditions 
arising from \ppp\ backgrounds. The classification of possible 
$Dp$-branes, branes at angles, intersecting branes and most importantly,
the theory residing on branes in the \ppp \ backgrounds and 
the corresponding \sugra\ solutions, along the lines of \cite{Kumar:2002ps,Alishahiha:2002rw},
deserve 
more detailed analysis.

{\large{\bf Acknowledgements}}

We would like to thank Keshav Dasgupta, Michal Fabinger, Jose Figueroa-O'Farrill,
Ori Ganor, Simeon Hellerman, Juan Maldacena
and Sergey Prokushkin for helpful discussions and comments.
The work of M. M. Sh-J. is supported in part by NSF grant
PHY-9870115 and in part by funds from the Stanford Institute for Theoretical 
Physics. The work of D. S. is supported by the Department of Energy,
Contract DE-AC03-76SF00515.

 \renewcommand{\theequation}{A-\arabic{equation}}
  \setcounter{equation}{0}  
  \section*{APPENDIX - Conventions}  

We briefly review our conventions in this appendix. We use the mostly plus metric.
Greek indices $\mu,\nu,...$ range over the curved (world) indices, while 
Latin indices
$a,b,...$ denote tangent space indices and $i,j$ label coordinates on the
space transverse to the light-cone directions.
The curved space Gamma matrices are defined via contraction with vierbeins
as usual, $\Gamma^\mu = e^\mu_a \Gamma^a$.

We may rewrite the two Majorana-Weyl spinors in ten dimensional type IIA and IIB theories
as a pair of Majorana spinors $\chi^\alpha, \alpha=1,2$,
subject to the chirality conditions appropriate to the theory,
\be \label{chirality}
\Gamma^{11} \: \chi^1 \: = \: + \: \chi^1 \ , \ \ \ \ \ \ \ \ \ \
\Gamma^{11} \: \chi^2 \: = \: \pm \: \chi^2 \ ,
\ee
where for the second spinor we choose $-$ for non-chiral type IIA
and $+$ for chiral type IIB theories, and treat the index $\alpha$ labeling the
spinor as an SL(2,$\mathbb{R}$) index. Where Pauli matrices appear, they
act on this auxiliary index, with
$\sigma_3$ acting analogously to the chirality operator in type IIA.
In what follows, we suppress the spinor index as well as
the auxiliary index.

Type II string theories contain two Majorana-Weyl gravitinos
$\psi_\mu^\alpha$, $\alpha=1,2$, which are of the same (opposite) chirality in IIB 
(IIA).
The supersymmetry variation of these gravitinos in string frame is
\be
\delta\psi_\mu \: = \: \hat{\mathcal{D}}_{\mu} 
\epsilon,
\ee
where the supercovariant derivative is defined as \cite{Cvetic,Alishahiha:2002nf}
\be
\hat{\mathcal{D}}_{\mu} =
\nabla_\mu
+\frac{1}{8} 
\sigma_3 \Gamma^{ab} H_{\mu ab} -
\frac{1}{16} e^{\phi}\left(\sigma_3 \Gamma^{ab} 
F_{ab}-\frac{1}{12}\Gamma^{abcd} F_{abcd}\right) 
\Gamma_\mu\ ,
\ee
for type IIA theory and 
\be
\hat{\mathcal{D}}_{\mu} =
\nabla_\mu
+\frac{1}{8} 
\sigma_3 \Gamma^{ab} H_{\mu ab}
+\frac{i}{8} e^{\phi}\left(\sigma_2 \Gamma^{a} 
\partial_a \chi-\frac{i}{6} 
\sigma_1\Gamma^{abc}F_{abc}+\frac{1}{240} 
\sigma_2 \Gamma^{abcde}F_{abcde}
\right) \Gamma_\mu\ ,
\ee
for type IIB, with the spin connection $\omega_\mu^{ab}$ appearing in the covariant derivative
$\nabla_\mu=(\partial_{\mu}+\frac{1}{4} \omega_\mu^{ab} 
\Gamma_{ab})$.
In these expressions, $\phi$ is the dilaton, $\chi$ is the axion, $H$ the three-form
field strength from the NS-NS sector, and the $F$'s represent the appropriate RR
field strengths.

For pure NS-NS backgrounds relevant to this paper, the dilatino variation for
both types IIA and IIB, is
\be
\delta \lambda \: = \:
\left(
\frac{1}{2} \Gamma^a ( \partial_a \phi )
\: - \: \frac{1}{4} \Gamma^{abc} H_{abc} \sigma_3
\right) \epsilon \ .
\ee


A convenient choice of basis for
$32 \times 32$ Dirac matrices, which we denote by $\Gamma^\mu$, 
can be written in terms of $16 \times 16$
matrices $\gamma^\mu$ such that
\be
\label{basis}
  \Gamma^{+} \: = \:
  i
  \left(
  \begin{matrix}
     0 & \sqrt{2}  \\
     0 & 0
  \end{matrix}
  \right) ,
  \ \ \ \
  \Gamma^{-} \: = \:
  i
  \left(
  \begin{matrix}
     0 & 0 \\
     \sqrt{2} & 0
  \end{matrix}
  \right) ,
  \ \ \ \
  \Gamma^{i} \: = \:
  \left(
  \begin{matrix}
     \gamma^i & 0 \\
     0 & - \gamma^i
  \end{matrix}
  \right) ,
  \ \ \ \
  \Gamma^{11} \: = \:
  \left(
  \begin{matrix}
     \gamma^{(8)} & 0 \\
     0 & -\gamma^{(8)}
  \end{matrix}
  \right) \ ,
\ee
and the $\gamma^i$ satisfying $\{\gamma^i,\gamma^j\}=2\delta_{ij}$ with $\delta_{ij}$
the metric on the transverse space.
Choosing a chiral basis for the $\gamma$'s, we have
$\gamma^{(8)}=diag(1_{8},-1_{8})$.
The above matrices satisfy
\be
\begin{split}
\label{properties}
(\Gamma^+)^\dagger \: = \: - \Gamma^- , \ \ \ \ \
(\Gamma^-)^\dagger \: &= \: - \Gamma^+ , \ \ \ \ \
(\Gamma^+)^2 = (\Gamma^-)^2 = 0 , \\
\left[ \Gamma^+ , \hplus \right] = 0 , \ \ \ \ \
\{ \Gamma^{11} , \Gamma^\pm \} \: = \: & 0 , \ \ \ \ \
\{ \Gamma^{11} , \Gamma^{i} \} \: = \: 0 , \ \ \ \ \
\left[ \Gamma^\pm , \Gamma^{ij} \right] \: = 0 \ ,
\end{split}
\ee
and
$\Gamma^\pm \Gamma^i ... \Gamma^j \Gamma^\pm = 0$ if the same signs appearing
on both sides.

We define light-cone coordinates
$x^{\pm} = (x^0 \pm x^9)/\sqrt{2}$ and likewise for the
light-like Gamma matrices $\Gamma^\pm=(\Gamma^0 \pm \Gamma^9) / \sqrt{2}$,
and also define antisymmetric products of $\gamma$ matrices with weight one,
$\gamma^{ab ... cd} \equiv \gamma^{[a} \gamma^b ... \gamma^c \gamma^{d]}$,
choosing for the $\gamma$ matrices a representation such that
\begin{subequations}
\begin{alignat}{3}
  \gamma^{12} \: &= \: &i \:
  diag(++--,++--,++--,++--) \ ,\\
  \gamma^{34} \: &= \: -&i \:
  diag(++--,--++,++--,--++) \ , \\
  \gamma^{56} \: &= \: &i \:
  diag(+-+-,+-+-,+-+-,+-+-) \ , \\
  \gamma^{78} \: &= \: -&i \:
  diag(+-+-,-+-+,-+-+,+-+-) \ ,
\end{alignat}
\end{subequations}
for which
\be
  \gamma^{12345678} \: \equiv \: \gamma^{(8)} \: = \:
  diag(++++,++++,----,----) \ .
\ee
The following combination appears in the paper,
which we write using the results of section \ref{ppp:hom} (in particular \eqref{foura}),
\be
\label{H:contraction}
\hplus \: \equiv \:
\Gamma^{ij} H_{+ij} \: = \:
\left(
\begin{matrix}
\gamma^{ij} H_{+ij} & 0 \\
0 & \gamma^{ij} H_{+ij}
\end{matrix}
\right) \ ,
\ee
and
\be \label{H:contraction:explicit}
\begin{split}
\gamma^{ij} H_{+ij} \: &= \:
2 \left( a_1 \gamma^{12} + a_2 \gamma^{34} + a_3 \gamma^{56} + a_4 \gamma^{78} \right) \\
&= \: 2 i \:
diag(c_1,c_2,-c_2,-c_1, c_3,c_4,-c_4,-c_3, c_5,c_6,-c_6,-c_5, c_7,c_8,-c_8,-c_7) \ ,
\end{split}
\ee
where we also define the constants
\begin{subequations}
\label{constants}
\begin{align}
c_1 \: &= \: a_1 - a_2 + a_3 - a_4 \ , \\
c_2 \: &= \: a_1 - a_2 - a_3 + a_4 \ , \\
c_3 \: &= \: a_1 + a_2 + a_3 + a_4 \ , \\
c_4 \: &= \: a_1 + a_2 - a_3 - a_4 \ , \\
c_5 \: &= \: a_1 - a_2 + a_3 + a_4 \ , \\
c_6 \: &= \: a_1 - a_2 - a_3 - a_4 \ , \\
c_7 \: &= \: a_1 + a_2 + a_3 - a_4 \ , \\
c_8 \: &= \: a_1 + a_2 - a_3 + a_4 \ .
\end{align}
\end{subequations}


\end{document}